\documentclass[showpacs,preprintnumbers,a4paper,amsmath,amssymb,nofootinbib,groupedaddress,superscriptaddress,showkeys]{revtex4}

\usepackage{graphicx}
\usepackage{fancybox}
\usepackage{dcolumn}
\usepackage{epsfig}
\usepackage[dvips]{color}
\usepackage{bm}

\allowdisplaybreaks[4]

\newcommand{\beq}{\begin{equation}}
\newcommand{\eeq}{\end{equation}}
\newcommand{\be}{\begin{eqnarray}}
\newcommand{\ee}{\end{eqnarray}}
\newcommand{\bav}{\begin{array}{cccc}}
\newcommand{\bea}{\begin{equation} \begin{array}{c}}
\newcommand{\eea}{ \end{array} \end{equation}}
\newcommand{\ea}{\end{array}}

\def\pmutau{P_{\nu_{\mu} \rightarrow \nu_{\tau}}}

\newcommand{\dms}{\mbox{$\Delta{m}^{2}_{21}$~}}
\newcommand{\dma}{\mbox{$\Delta{m}^{2}_{31}$~}}

\newcommand{\epse}{\mbox{$\epsilon_e$}}
\newcommand{\epsmu}{\mbox{$\epsilon_\mu$}}
\newcommand{\epstau}{\mbox{$\epsilon_\tau$}}
\def\gs{\mathrel{
   \rlap{\raise 0.511ex \hbox{$>$}}{\lower 0.511ex \hbox{$\sim$}}}}
\def\ls{\mathrel{
   \rlap{\raise 0.511ex \hbox{$<$}}{\lower 0.511ex \hbox{$\sim$}}}}

\begin{document}

\title{
Testing non-unitarity of neutrino mixing matrices at  neutrino
factories
}

\author{Srubabati Goswami}
\email{sruba@mri.ernet.in}
\affiliation{Harish--Chandra Research Institute, Chhatnag Road,\\
Jhunsi, Allahabad 211 019, India
}

\author{Toshihiko Ota}
\email{Toshihiko.Ota@physik.uni-wuerzburg.de}
\affiliation{Institut f\"{u}r Theoretische Physik und Astrophysik
Universit\"{a}t W\"{u}rzburg,\\
Am Hubland
97074 W\"{u}rzburg, Germany
}

\date{\today}

\thispagestyle{empty}

\begin{abstract}
In this paper we explore  
the effect of non-unitary neutrino mixing  on
neutrino oscillation probabilities both in vacuum and matter. 
In particular, we consider the
$\nu_\mu \rightarrow \nu_\tau$ channel and using a Neutrino Factory
as the source for $\nu_\mu$'s discuss the constraints that 
can be obtained on the moduli and phases of the parameters characterizing 
the violation of unitarity.  
We  point out  how the new CP violation 
phases present in the case where the non-unitary mixings 
give rise to
spurious ``degenerate'' solutions in the parameter space and 
discuss how the true solutions can be extricated by combining 
measurements at several baselines. 
\end{abstract}
\maketitle

\section{Introduction}

There is a phenomenal increase in our knowledge of neutrino properties in
the past few years coming from neutrino oscillation data  from solar, 
atmospheric,
accelerator and reactor neutrino experiments.
For three neutrino flavours, there are nine parameters characterizing
the light neutrino mass matrix, the three masses, three mixing angles and
three CP phases.
Neutrino oscillation data determines the
best-fit values and the $3\sigma$ ranges of the mass squared differences
and mixing angles as \cite{global} 

\begin{itemize}
\item Combined analysis of solar and KamLAND reactor neutrino data gives 
the best-fit values and 3$\sigma$ ranges of mass and mixing parameters as 
$\dms \equiv m_2^2 - m_1^2 = 7.9^{+1.0}_{-0.8} \cdot 10^{-5}$ eV$^2$ 
and $\sin^2 \theta_{12} = 0.31^{+0.09}_{-0.08}$. 
The solar data implies $\Delta m^2_{21} > 0$ .  

\item Global analysis of atmospheric neutrino data from SuperKamiokande 
and data from accelerator experiments K2K and MINOS gives 
$\mid{\Delta m^2_{31}}\mid \equiv |m_3^2 - m_1^2| =2.5^{+0.7}_{-0.6}
      \cdot 10^{-3}$ eV$^2$ 
and
$\sin^2\theta_{23} = 0.5^{+0.38}_{ -0.16}$. 

\item 
The value of the third leptonic mixing angle  $\theta_{13}$ is not yet known
and at present it is bounded to be $\sin^2\theta_{13} < 0.05$ leaving open the
possibility of very small or zero value for this.
\end{itemize} 
This tremendous progress has initiated the precision era of 
neutrino physics, and experiments are planned and proposed to 
further increase the precision of the known neutrino parameters 
and to pin-down the value of the mixing angle $\theta_{13}$ 
and determine the sign of $\Delta m^2_{31}$ ($\text{sign}[\Delta m^2_{31}]$) 
\footnote{
Usually $\Delta m^2_{31} >0$  and 
$m_3^2 \simeq \dma \gg m_2^2 \simeq \dms \gg m_1^2$
is referred to as 
normal hierarchy (NH), 
and $\Delta m^2_{31} <0$ and 
$m_2^2 \simeq |\dma|+\dms > m_1^2 \simeq |\dma| \gg m_3^2$
as inverted hierarchy (IH).
The three neutrinos can also be quasi-degenerate
with $m_3^2 \simeq m_2^2 \simeq m_1^2 \equiv m_0^2 \gg |\dma|$ 
in which there is no hierarchy. However, one can still ask 
what the sign of  
$\Delta m^2_{31}$ is. }.
                           
A non-zero value of $\theta_{13}$ is intimately related
to the possibility of observation of CP phase in the lepton sector.
A large value of $\theta_{13}$ would also enable one to determine the 
$\text{sign}[\Delta m^2_{31}]$ through observation of large
matter effects for neutrinos propagating through 
earth \cite{pb,us2,usnew,oscillogram}. 
If $\theta_{13}$ is relatively large, $\sin^2 2\theta_{13} \gs 0.01$, 
then the answers to these questions may be obtained from 
superbeam \cite{Huber:2002mx,superbeam} and future atmospheric neutrino
experiments 
\cite{usnew,us2,futureatm1,futureatm2,futureatm3,futureatm4,futureatm5}. 
However, if Nature selects $\theta_{13}$ to be smaller than this, 
then one has to go to either $\beta$-beam or neutrino factory 
experiments.  
The R\&D for both are actively pursued \cite{Albright:2004iw,ISSreport}. 
Future facilities also have the potential to discover new 
physics \cite{sanjib1,sanjib2,ota,KLOS,Minakata1,Minakata2}. 

The best-fit values of masses and mixing angles quoted above are obtained 
assuming the neutrino mixing matrix 
(Pontecorvo-Maki-Nakagawa-Sakata (PMNS) matrix)
to be unitary. 
However, for models with heavy fermionic fields, the deviation of 
the leptonic mixing matrix from unitarity is a generic feature 
\cite{ll,antusch,gavela}.
A typical example is the type-I seesaw mechanism \cite{seesaw1,seesaw2,seesaw3,seesaw4,seesaw5} which provides 
a natural framework of generating small neutrino masses. 
This requires introduction of one or more heavy
right handed singlet neutrino field(s). 
Although the full mixing matrix at the high scale is expected to be unitary
in these cases, the mixing matrix relevant for  low energy phenomenology
is not unitary as the production of the heavy particles are
kinematically forbidden.
However the violation from unitarity in the canonical Type-I 
seesaw mechanism
is found to be very small if the mass scale of the heavy neutrinos 
are of the order of the GUT scale $\sim 10^{16}$ GeV 
and the heavy neutrinos decouple and do not influence the physics 
at low scale. 
However non-minimal seesaw models have been constructed with 
heavy neutrinos of mass ${\cal{O}}$(1) TeV, invoking symmetry 
arguments to suppress the seesaw term 
\cite{tevseesaw1,tevseesaw2,tevseesaw3,tevseesawOkamura}.  
 Such models can give rise to 
significant light-heavy mixing and deviation from unitarity. 
The TeV scale seesaw models are interesting as these can have 
signatures in the Large Hadron Colliders (LHC) in the near future
\cite{lhc1,lhc2,lhc3}. 
Also successful leptogenesis can be generated if the heavy Majorana
neutrinos are quasi-degenerate \cite{underwood1,underwood2,lepto-xing}.
There are also models with heavy neutral (gauge singlets) which can give 
large light-heavy mixings \cite{ravivalle,generic1,generic2}. 
In the $R$-parity violating supersymmetric models,
neutrinos can also mix with neutralinos \cite{RpV}.
Since deviation from
unitarity is due to the physics at the high scale, a measurement of
them at the low scale can  serve a window to the physics at high energy.
Hence it is important to probe if the future precision neutrino experiments
can
give any indication towards the non-unitary nature of neutrino mixing matrix.
In this paper we address  this question. 

The non-unitary nature of the neutrino mixing matrix due to  mixing
with fields  heavier than $M_Z/2$
can manifest itself in tree level process like $\pi \rightarrow \mu \nu$,
$Z \rightarrow \bar{\nu} {\nu}$, $W \rightarrow l \nu$ or in flavour
violating
rare charged lepton decays like $\mu \rightarrow e \gamma$, $\tau \rightarrow
\mu \gamma$ etc., which proceed via one-loop processes and hence 
can be constrained from low energy electroweak data 
\cite{ll,generic1,generic2,antusch,lh1,lh2,lh3,lh4,lh5,kagan}. 
Non-unitarity of neutrino mixing matrices can also affect the 
neutrino oscillation probabilities 
\cite{nonu-osc,Czakon:2001em,delAguila:2002sx,Bekman:2002zk,Holeczek:2007kk,yasuda}. 
In this paper, we concentrate on the effect of non-unitarity on
neutrino oscillation probabilities and the possibility of probing this in
neutrino factories. 
We show that the effect of non-unitarity can be
more pronounced  in the appearance channel than in the survival channel.
In particular, we look into the effect of deviation from non-unitarity in
the $\nu_\mu$-$\nu_\tau$ channel since the present constraint on the
non-unitarity parameter in this channel is much weaker than the constraint on
the $\nu_e$-$\nu_\mu$ channel.
We consider $\nu_\tau$ detectors like OPERA \cite{opera}
or ICARUS \cite{icarus}
detectors for CERN Neutrinos to Gran Sasso (CNGS) 
$\nu_\mu \rightarrow \nu_\tau$
oscillation  search programme
and discuss the possibility of constraining the
moduli and phases parametrising the unitarity violation.
These phases characterizing the non-unitarity constitute a new source for 
CP violation which can be present even in the limit of 
$\theta_{13}\rightarrow 0$. 
We also discuss the matter effects in the presence of non-unitarity and
show that for a non-unitary mixing matrix, 
matter effect can manifest itself
even in the limit of the third leptonic mixing angle 
$\theta_{13} \rightarrow 0$ and 
in the One Mass Scale Dominance (OMSD) limit of
$\Delta m^2_{21}/\Delta m_{31}^{2} \rightarrow 0$.
There is some overlap of our work with Ref.~\cite{yasuda} 
who have also constrained 
non-unitarity violation using 
the  $\nu_\mu \rightarrow \nu_\tau$ channel. 
However, we consider the possibility
of combining several baselines, reducing the degeneracy of parameter
space.
To distinguish the non-unitarity signature 
with that of non-standard interactions, the  
combination of the baselines is useful.
When two or more observations 
suggest the same parameter region for scenarios with 
non-unitary lepton mixing matrix, 
there can be stronger implications to
determine the origin of the signal 
beyond the standard oscillation scenario.

The plan of the paper is as follows. In the next section
discuss the parametrization that we use for non-unitary mixing matrices and
present the current constraints on unitarity violation.
In section III
we give simplified
expressions for the oscillation probabilities  in vacuum
and matter assuming the
mixing matrix to be non-unitary.
In section IV  we discuss the
degeneracies in the oscillation probabilities. In section V 
we give our numerical results on the allowed regions of the 
parameter space in the model with the non-unitary PMNS matrix.  
We conclude in section VI.

\section{Non-unitary mixing matrices and current constraints } 

Since non-unitarity of mixing matrices is a generic feature
of theories with heavy neutrinos we consider a picture with three
light and one heavy neutrino. In this case the full 4$\times$4
mixing matrix is
unitary but the 3$\times$3 light neutrino submatrix is non-unitary.
A $4\times4$ unitary matrix can be parametrized by 6 angles 
$\theta_{12,13,14,23,24 ,34}$ and three phases $\delta_{13, 24,34}$.
If the neutrinos are Majorana in nature then 
three additional phases can be present. 
We parametrize the 4$\times$4 unitary matrix in the usual way  in terms
of the
rotation matrices $R_{ij}$
\be
\mathcal{U} 
=  \tilde{R}_{34} \tilde{R}_{24} R_{14} R_{23} 
\tilde{R}_{13} R_{12} P
\label{fullu}
\ee
where the $R_{ij}$ represent rotations in $ij$ generation space,
for instance:
\be
\tilde{R}_{34} =
\left(
\bav
1 & 0 & 0 & 0 \\
0 & 1 & 0 & 0 \\
0 & 0 & c_{34} & s_{34} {\rm e}^{-{\rm i} \delta_{34}} \\
0 & 0 & -s_{34}{\rm e}^{{\rm i} \delta_{34}} & c_{34}
\ea
\right)~\mbox{ or }~
R_{14} =
\left(
\bav
c_{14} & 0 & 0 & s_{14} \\
0 & 1 & 0 & 0 \\
0 & 0 & 1 & 0 \\
-s_{14}& 0 & 0 & c_{14}
\ea
\right)~,
\ee
with the usual notation $s_{ij} = \sin \theta_{ij}$ and
$c_{ij} = \cos \theta_{ij}$.
The symbol tilde means the mixing matrix including the CP phase.
The diagonal matrix $P$ contains the three
Majorana phases, which we denote $\alpha, \beta$ and $\gamma$:
\be \label{eq:P}
P = {\rm diag} \left(1, {\rm e}^{-{\rm i} \alpha/2}, {\rm e}^{-{\rm i}(\beta/2 - \delta_{13})},
{\rm e}^{-{\rm i}(\gamma/2 - \delta_{34})} \right)~.
\ee
Since the Majorana phases are not important for oscillation studies 
henceforth we will omit the matrix $P$.  

Assuming the mixing of the fourth heavy state
to be small, the above equation can be expanded in terms of small
parameters $\epsilon_e$, $
\epsilon_\mu$ and $\epsilon_\tau$ characterizing the
14, 24 and 34 rotations respectively\footnote{We use
$\cos\theta_{ij} = \cos\theta_{ji} \simeq 1 - \epsilon_\alpha^2/2$ 
and $\sin\theta_{ij} = -\sin\theta_{ji} \simeq \epsilon_\alpha$,  
where
$\alpha$ is the corresponding index, $e$, $\mu$ or $\tau$.}.
With this simplification Eq.~\eqref{fullu} can be expressed as 
\beq
\mathcal{U} = \left(\begin{array}{cccc}
& &  & \epsilon_e  \\
& W  & & {\rm e}^{-{\rm i}\delta_{24}}\epsilon_\mu \\
& &  & {\rm e}^{-{\rm i}\delta_{34}}\epsilon_\tau \\
\mathcal{U}_{s1}  
& \mathcal{U}_{s2} 
& \mathcal{U}_{s3} 
& 1-\frac{1}{2}(\epsilon_e^2 +\epsilon_\mu^2 + \epsilon_\tau^2) \\
\end{array} \right)
\label{unitary}
\eeq
where $W$ is the $3 \times 3$ non-unitary mixing matrix. 
This can be written as,
\beq
W = \left(\begin{array}{ccc}
U_{e1} (1 - \epsilon_e^2/2) 
& U_{e2} (1 - \epsilon_e^2/2) 
& U_{e3} (1 - \epsilon_e^2/2) \\ [0.2cm]
U_{\mu1 } (1 - \epsilon_\mu^2/2) 
& U_{\mu2 } (1 - \epsilon_\mu^2/2) 
& U_{\mu3 } (1 - \epsilon_\mu^2/2) \\
- {\rm e}^{-{\rm i}\delta_{24}} \epsilon_\mu \epsilon_e U_{e1} 
& - {\rm e}^{-{\rm i}\delta_{24}} \epsilon_\mu \epsilon_e U_{e2} 
& - {\rm e}^{-{\rm i}\delta_{24}}  \epsilon_\mu \epsilon_e U_{e3} \\ [0.2cm]
U_{\tau1 } (1 - \epsilon_\tau^2/2) 
& U_{\tau2 } (1 - \epsilon_\tau^2/2) 
& U_{\tau3 } (1 - \epsilon_\tau^2/2) \\
- {\rm e}^{-{\rm i}\delta_{34}} \epsilon_e \epsilon_\tau U_{e1} 
 & -{\rm e}^{-{\rm i}\delta_{34}} \epsilon_e \epsilon_\tau U_{e2} 
 & -{\rm e}^{-{\rm i}\delta_{34}} \epsilon_e \epsilon_\tau U_{e3} \\
- {\rm e}^{{\rm i} \phi} \epsilon_\mu \epsilon_\tau U_{\mu1}
 & -{\rm e}^{{\rm i} \phi} \epsilon_\mu \epsilon_\tau U_{\mu2} 
 & -{\rm e}^{{\rm i} \phi}\epsilon_\mu \epsilon_\tau U_{\mu3} 
\end{array} \right)
\eeq
were $\phi = \delta_{24} - \delta_{34}$, 
$\mathcal{U}_{sk} = - \epse U_{ek} 
-  {\rm e}^{{\rm i} \delta_{24}} \epsmu U_{\mu k} 
-  {\rm e}^{{\rm i} \delta_{34}} \epstau U_{\tau k}$,
and the $3\times 3$ matrix $U_{\alpha i}$ 
with $\alpha=e,\mu,\tau$ and $i=1,2,3$ is defined and parameterized as
the usual unitary PMNS matrix for three generations.

Bound on the  moduli of the unitarity violation parameters can come from 
electroweak processes and  from neutrino oscillations. 
The bounds obtained from present neutrino oscillation experiments are 
weaker than those obtained from electroweak decays \cite{antusch}. 
Constraint on 
$\sum_{i=1}^{3} W_{\alpha i} W_{\beta i}^{*}$
comes from rare decays of charged leptons 
$\l_\alpha \rightarrow\l_\beta \gamma$
\cite{ll,generic1,generic2,underwood1,underwood2,antusch}.
Whereas $\sum_{i=1}^{3} |W_{\alpha i}|^{2}$ 
can be constrained from processes like 
$W \rightarrow l \nu$, $Z \rightarrow \nu \bar{\nu}$.  
Constraints on the diagonal elements of the non-unitary matrix 
can also come from tests for lepton universality 
\cite{generic1,generic2,antusch,ll}. 
At present there is strict constraint on 
light-heavy mixing  in the 
$e$-$\mu$ sector coming from non-observation of the decay 
$\mu \rightarrow e \gamma$. 
For non-unitarity induced through heavy right handed neutrinos  
the bound quoted in Ref.~\cite{underwood1,underwood2} is
\be
\left|\sum_{i=1}^{3} W_{e i} W_{\mu i}^{*} \right|
\equiv \epsilon_e \epsilon_\mu \ls  1.2 \cdot 10^{-4} 
\ee
The bound on the $\mu$-$\tau$ sector is much weaker 
\be
\left|\sum_{i=1}^{3} W_{\mu i} W_{\tau i}^{*} \right|
 \equiv \epsilon_\mu \epsilon_\tau \ls 2\cdot 10^{-2} 
\ee
The $\epsilon_\alpha$'s are  also constrained by electroweak measurements 
individually as \cite{lhc2,kagan}
\be 
\epsilon_{e}^{2}< 0.012, 
~~\epsilon_{\mu}^{2}< 0.0096, 
~~\epsilon_{\tau}^{2} < 0.016.
\ee 

\section{Calculation of Oscillation Probabilities} 

\subsection{Oscillation Probability in Vacuum} 

The most general expression of survival/oscillation probability for
$\nu_{\alpha} \rightarrow \nu_{\beta}$ in vacuum without assuming unitarity of 
mixing matrices is \cite{Czakon:2001em}
\begin{align}
 P_{\nu_{\alpha}\rightarrow\nu_{\beta}}
 =&
  \frac{1}{
  N_\alpha
  N_\beta}
 \Biggl\{
 \left|
 \sum_{i=1}^{\mathrm{light}}
 {{W}_{\beta i}} {W^{*}_{\alpha i}}
 \right|^{2}
 -
 4
 \sum_{i<j}^{\mathrm{light}}
 R_{\alpha \beta}^{ij}
 \sin^{2} \frac{(m_{j}^{2}-m_{i}^{2})L}{4E}
 -
 2
 \sum_{i<j}^{\mathrm{light}}
 I_{\alpha \beta}^{ij}
 \sin \frac{(m_{j}^{2}-m_{i}^{2})L}{2E}
 \Biggr\},
\end{align}
where 
$N_{\alpha} = \sum_{i=1}^{\mathrm{light} } |W_{\alpha i}|^2$;
$ R_{\alpha \beta}^{ij} = {\rm Re}[W_{\beta i} W^{*}_{\alpha i}
W^{*}_{\beta j} W_{\alpha j}]$; 
$ I_{\alpha \beta}^{ij} = {\rm Im}[W_{\beta i} W^{*}_{\alpha i} W^{*}_{\beta j}
W_{\alpha j}]$, 
and the sum of 
the mass eigenstate index is taken with the states which concern with
the neutrino propagation (which is mentioned as ``light'' here).
Although we consider a $4 \times 4$ mixing matrix (for 
the three light mass eigenstates and one heavy one) 
in the previous section 
and in the rest of the paper, the above expression for probability  
can be applied to the more general case  
where $W$ is the part
of the larger unitary matrix than $4 \times 4$. 

If we concentrate on baselines and energies such that the
OMSD approximation can be employed, then the
terms
containing $\Delta m_{21}^{2} L / (4E)$ can be neglected and the
expression simplifies to
\begin{align}
 P_{\nu_{\alpha}\rightarrow\nu_{\beta}}
  =&
\frac{1}{ N_\alpha N_\beta}
     \Biggl\{
 \left|        
 \sum_{i=1}^{3}
  W_{\beta i} W^{*}_{\alpha i}
 \right|^{2}
 -
  4
 [R_{\alpha \beta}^{13} + R_{\alpha \beta}^{23}]
     \sin^{2} \frac{\Delta m_{31}^{2} L}{4E}
      -
       2
       [I_{\alpha \beta}^{13} + I_{\alpha \beta}^{23}]
          \sin \frac{\Delta m_{31}^{2}L}{2E}
           \Biggr\}.
\end{align}           
As mentioned above,
there is already strong constraint on the combination of the parameters 
$\epsilon_{e} \epsilon_{\mu}$.
Therefore we assume $\epsilon_e = 0$ throughout this article.
With this assumption, 
the deviation of unitarity can occur in the 
$\nu_{\mu}\rightarrow \nu_{\mu}$ , $\nu_{\mu}\rightarrow \nu_{\tau}$ 
and
$\nu_{\tau}\rightarrow \nu_{\tau}$ channel\footnote{%
Alternatively one can study the 
violation of unitarity in both   
$\nu_{e}\rightarrow\nu_{\tau}$ channel 
and 
$\nu_{\mu}\rightarrow \nu_{\tau}$ channel \cite{yasuda}.}.
In the limit of $\theta_{13} \rightarrow 0$ and 
$\Delta m^2_{21}/\Delta m_{31}^{2} \rightarrow 0$, 
the survival probability $P_{\nu_{\mu}\rightarrow \nu_{\mu}}$ 
can be expressed as 
\begin{equation} 
  P_{\nu_{\mu}\rightarrow \nu_{\mu}} 
   = 1 - \sin^{2} 2 \theta_{23} 
  \sin^2 \frac{\Delta m_{31}^2 L}{4E}
  + \mathcal{O}(\epsilon^{3}).
\end{equation} 
From this equation, we see that the second order of 
the non-unitary effects in each term 
cancel out with the normalization factor 
$1/N_{\mu}^{2}$.
In the $\nu_{\tau} \rightarrow \nu_{\tau}$ channel,
the non-unitary effect comes as a small correction to the standard
oscillation term and the standard oscillation term dominates.  
On the other hand,
the oscillation probability for $\nu_{\mu} \rightarrow \nu_{\tau}$ is
approximated as  
\begin{align}
P_{\nu_{\mu} \rightarrow \nu_{\tau}}
 =&
 \left\{
 \epsilon_{\mu}^{2} \epsilon_{\tau}^{2}
 +
 \mathcal{O}(\epsilon^{5})
 \right\} \nonumber \\
 &+
 \sin 2 \theta_{23}
 \left\{
 \sin 2 \theta_{23} 
 +
 2 \epsilon_{\mu} \epsilon_{\tau} \cos 2 \theta_{23} \cos\phi
 +
 \mathcal{O} (\epsilon^{3})
 \right\}
 \sin^{2} \frac{\Delta m_{31}^{2} L}{4E} \nonumber \\
 &+
 \left\{
 \epsilon_{\mu} \epsilon_{\tau}
 \sin \phi
 \sin 2\theta_{23}
 +
 \mathcal{O}(\epsilon^{3})
 \right\}
 \sin \frac{\Delta m_{31}^{2} L }{2E} \nonumber \\
&
 +
 \mathcal{O}(s_{13})
 +
 \mathcal{O}(\Delta m_{21}^{2}/\Delta m_{31}^{2}).
\label{eq:Pmutau}
\end{align}
The term with $\sin\phi$ takes a different energy dependence from 
the standard oscillation term.
Therefore, we can expect that this can be distinguished
from the standard oscillation signals.
The term of $\mathcal{O}(\epsilon^{2})$ in the standard
oscillation term 
($\sin^{2} \Delta m_{31}^{2} L/(4E)$ term) cannot be important
because it is always smaller enough than 
the standard contribution $\sin^{2} 2 \theta_{23}$.
Assuming $L=130$ km, $E = 50$ GeV, and 
$\epsilon_{\mu}\epsilon_{\tau} = 10^{-2}$, 
the order of each term is calculated to be
\begin{align}
& \text{standard oscillation term: } 
 \sin^{2}  \frac{\Delta m_{31}^{2} L}{4E}
 \sim 6.8 \cdot 10^{-5}, \\
& \text{$\sin\phi$ term: } 
 \epsilon_{\mu} \epsilon_{\tau}
 \sin  \frac{\Delta m_{31}^{2} L}{2E}
 \sim 1.7 \cdot 10^{-4}, \\
& \text{zero-distance term: }
 \epsilon_{\mu}^{2} \epsilon_{\tau}^{2}
 = 
 10^{-4},
\end{align}
and the three terms in Eq.~\eqref{eq:Pmutau} 
are thus of the same order of magnitude and this channel 
provides a better option for probing violation of unitarity.

The noteworthy feature of the above equation is 
the zero-distance term $\epsilon_{\mu}^{2} \epsilon_{\tau}^{2}$. 
Consequently for a near detector one gets, 
\begin{equation}
 P_{\nu_{\mu} \rightarrow \nu_{\tau}}^{\text{near}}
 =
 \epsilon_{\mu}^{2} \epsilon_{\tau}^{2}.
\label{eq:oscP-near}
\end{equation}
It is actually very small.
However, there are two positive aspects: 
(i) a huge number of neutrinos comes into the near detector
(ii) the background for this process,
i.e., the standard oscillation events,
is highly suppressed.

\subsection{Oscillation Probability in Matter} 

When we introduce the non-unitary PMNS matrix, 
neutrinos obtain the additional matter effect mediated by 
neutral current 
\cite{delAguila:2002sx,Bekman:2002zk,Holeczek:2007kk}\footnote{%
The non-standard matter effect mediated by neutral current
interactions was also discussed in Ref.~\cite{NCMatter}.}.
For non-unitary mixing, the $\nu_\mu \rightarrow \nu_\tau$ oscillation 
probability in matter of constant density  
in  the simplifying approximation of $\theta_{13} \rightarrow 0$ and 
$\Delta m^2_{21}/\Delta m_{31}^{2} \rightarrow 0$ can be expressed as,
\begin{align}
 P_{\nu_{\mu} \rightarrow \nu_{\tau}}
 =&
 \sin 2\theta_{23}
 \left(
 \sin 2 \theta_{23}
 +
 2 \epsilon_{\mu} \epsilon_{\tau} \cos 2\theta_{23} \cos\phi
 \right)
 \sin^{2} \frac{\Delta m_{31}^{2} L}{4E}
 +
 \epsilon_{\mu} \epsilon_{\tau}
 \sin \phi
 \sin 2\theta_{23}
\sin \frac{\Delta m_{31}^{2} L }{2E} \nonumber \\
&-
 \epsilon_{\mu} \epsilon_{\tau}
 \left( \frac{a_{\rm NC} L}{2E} \right)
 \sin^{3} 2 \theta_{23} \cos\phi
 \sin \frac{\Delta m_{31}^{2} L}{2E} 
 -
 4 \epsilon_{\mu} \epsilon_{\tau}
 \left(\frac{a_{\rm NC}}{\Delta m_{31}^{2}}\right)
 \sin 2 \theta_{23} \cos^{2} 2 \theta_{23} \cos\phi
 \sin^{2} \frac{\Delta m_{31}^{2} L}{4E}
 \nonumber \\
 &
 -2
 \left(\frac{a_{\rm NC}}{\Delta m_{31}^{2}} \right)
 \sin^{2} 2\theta_{23} \cos 2 \theta_{23}
 (\epsilon_{\mu}^{2} -\epsilon_{\tau}^{2})
 \sin^{2} \frac{\Delta m_{31}^{2} L}{4E} 
 +
 \left(
 \frac{a_{\rm NC} L}{4E}
 \right)
 \sin^{2} \theta_{23} \cos 2 \theta_{23}
 (\epsilon_{\mu}^{2} - \epsilon_{\tau}^{2})
 \sin \frac{\Delta m_{31}^{2} L}{2E} \nonumber \\
 &+
 \mathcal{O}(\epsilon^{3})
 +
 \mathcal{O}(s_{13})
 +
 \mathcal{O}(\Delta m_{21}^{2}/\Delta m_{31}^{2}),
\end{align}
where $a_{\rm NC}$ is the matter effect mediated 
by neutral current interaction.
This is consistent with the result shown in Ref.~\cite{Holeczek:2007kk}
though the procedures used are somewhat different.
Since $\theta_{23} \simeq \pi/4$, we can omit the terms which
proportional to $\cos 2\theta_{23}$, and
finally, it is reduced to
\begin{align}
 P_{\nu_{\mu} \rightarrow \nu_{\tau}}
  =&
 \sin^{2} 2\theta_{23}
 \sin^{2} \frac{\Delta m_{31}^{2} L}{4E}
 +
 \epsilon_{\mu} \epsilon_{\tau}
 \sin 2\theta_{23}
 \sin \phi
 \sin \frac{\Delta m_{31}^{2} L }{2E} 
 -
 \epsilon_{\mu} \epsilon_{\tau}
 \left( \frac{a_{\rm NC} L}{2E} \right)
 \sin^{3} 2 \theta_{23}
 \cos\phi
 \sin \frac{\Delta m_{31}^{2} L}{2E}.
\label{pmutaumat2}
\end{align}
This formula can nicely explain the numerical result which will be shown
in the following sections.
We have an additional term in comparison with Eq.~\eqref{eq:Pmutau},
which depends on $\cos\phi$ differing from the vacuum term.
This is the key feature to resolve the degeneracies which will be
explained in the next section.
The details of the derivation are described in Appendix.

\section{Degeneracies} 

From the expression 
Eq.~\eqref{eq:Pmutau} for the oscillation probability  
$\pmutau$ in vacuum, 
we see that this is invariant under the following transformations:  
\begin{enumerate}
 \item $\theta_{23}$ (octant) degeneracy:
       $P_{\nu_{\mu} \rightarrow \nu_{\tau}}
       (\theta_{23})
       =
       P_{\nu_{\mu} \rightarrow \nu_{\tau}}(\pi/2 - \theta_{23})$,

 \item $\text{sign}[\Delta m_{31}^{2}]$-$\phi$ degeneracy:
       $P_{\nu_{\mu} \rightarrow \nu_{\tau}}
       (\Delta m_{31}^{2}>0, \phi)
       =
       P_{\nu_{\mu} \rightarrow \nu_{\tau}}
       (\Delta m_{31}^{2}<0, -\phi)$,

 \item $\phi$-$(\pi-\phi)$ degeneracy:
       $P_{\nu_{\mu} \rightarrow \nu_{\tau}}
       (\phi)
       = P_{\nu_{\mu} \rightarrow \nu_{\tau}}
       (\pi-\phi)$,

 \item $(\epsilon_{\mu} \epsilon_{\tau})$-$\phi$ correlation 
       (quasi-degeneracy):
       $P_{\nu_{\mu} \rightarrow \nu_{\tau}}
       \left( (\epsilon_{\mu} \epsilon_{\tau}),\phi \right)
       = P_{\nu_{\mu} \rightarrow \nu_{\tau}}
       \left( (\epsilon_{\mu} \epsilon_{\tau})',\phi' \right)$.
\end{enumerate}
Here, the values of oscillation parameters which are not explicitly shown 
are taken to be the same on both the sides of the equations.
These can give rise to degeneracies in the 
$(\epsilon_{\mu} \epsilon_{\tau})$-$\phi$ plane even in the 
limit $\theta_{13} \rightarrow 0$. 
Below we discuss these degeneracies. If $\theta_{13}$ is non-zero then 
the additional degeneracies due to $\delta_{\rm CP}$ can also be there. 
But this will not give rise to any additional degenerate solutions in the
$(\epsilon_{\mu} \epsilon_{\tau})$-$\phi$ plane.  
Note that in addition to the degeneracies  
$\epsmu \epstau$ and $\phi$ occur in a correlated fashion 
in the oscillation probability shown in Eq.~\eqref{eq:Pmutau}. 
Hence the uncertainty in determination of one of these parameters 
can affect that of the other even for the same hierarchy. 
When we assume the maximal mixing for $\theta_{23}$,
the $\theta_{23}$ octant degeneracy is not present.

The expression Eq.~\eqref{pmutaumat2} breaks some of the degeneracies. 
Because of the presence of the $\cos\phi$ term, induced by matter effect,
$\text{sign}[\Delta m_{31}^{2}]$-$\phi$ and 
$\phi$-$(\pi-\phi)$ degeneracies can be resolved 
if we can see this term.
To do so, we have to go to the long baseline because
the term is simply proportional to the baseline length.
However, in the long baseline region, 
the standard oscillation term can be order one, and 
the tiny non-unitarity effect could be easily absorbed 
by the standard oscillation term.
The significance of the non-unitary matter effect
should be checked numerically. 
The $(\epsmu \epstau)$-$\phi$
correlation is present in the oscillation probability in matter 
as well. 

We can illustrate the occurrence of degeneracies due to the 
invariance listed above by using the equi-probability 
plots \cite{Autiero:2003fu}. 
Here, the standard oscillation parameters are fixed as, 
\begin{gather}
\sin^{2}\theta_{12}=0.31, 
\qquad
\sin^{2} 2\theta_{13} = 10^{-2}, 
\qquad 
\delta_{\rm CP} = 0, \nonumber \\
\left|\Delta m_{31}^{2}\right|
= 2.5 \cdot 10^{-3} \text{ [eV$^{2}$]},
\qquad
\Delta m_{21}^{2}
= 7.9 \cdot 10^{-5} \text{ [eV$^{2}$]},
\label{eq:standardosc-ref}
\end{gather}
and $\theta_{23}$ and the sign of the
atmospheric mass square difference will be given later.
For the non-unitary parameters, 
we adopt 
\begin{align}
(\epsilon_{\mu} \epsilon_{\tau})^{\text{true}} = 10^{-2}, 
\qquad 
\phi^{\text{true}}=\pi/4
\label{eq:non-uni-ref}
\end{align}
as the reference values throughout this paper\footnote{%
More precisely, we take 
$\epsilon_{\mu}^{\text{true}} = \epsilon_{\tau}^{\text{true}} = 0.1$
in our numerical calculations.
This allocation does not affect the results
since the leading contribution of the non-unitarity 
always appears as the combination $\epsilon_{\mu}\epsilon_{\tau}$.
}.
The equi-probability curves shown in the following
mean that the condition
\begin{align}
 P_{\nu_{\mu} \rightarrow \nu_{\tau}}
 \left(
 (\epsilon_{\mu}\epsilon_{\tau})^{\text{fit}}, \phi^{\text{fit}}
 \right)
 =
 P_{\nu_{\mu} \rightarrow \nu_{\tau}}
 \left(
 (\epsilon_{\mu}\epsilon_{\tau})^{\text{true}},
 \phi^{\text{true}}
 \right),
\label{eq:equiP-condition}
\end{align}
is fulfilled on each curve.

\begin{figure}[tbh]
\unitlength=1cm
\begin{picture}(18,6)
\put(0,0){\includegraphics[width=6cm]{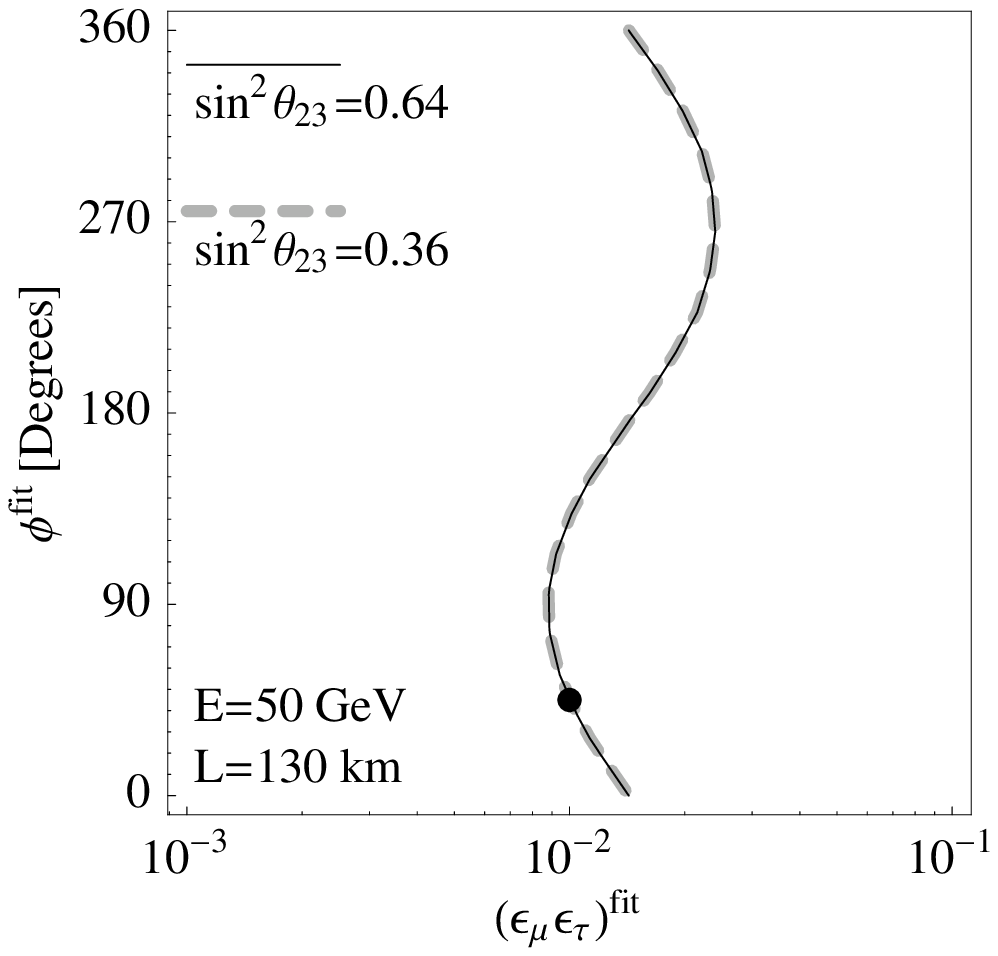}}
\put(6,0){\includegraphics[width=6cm]{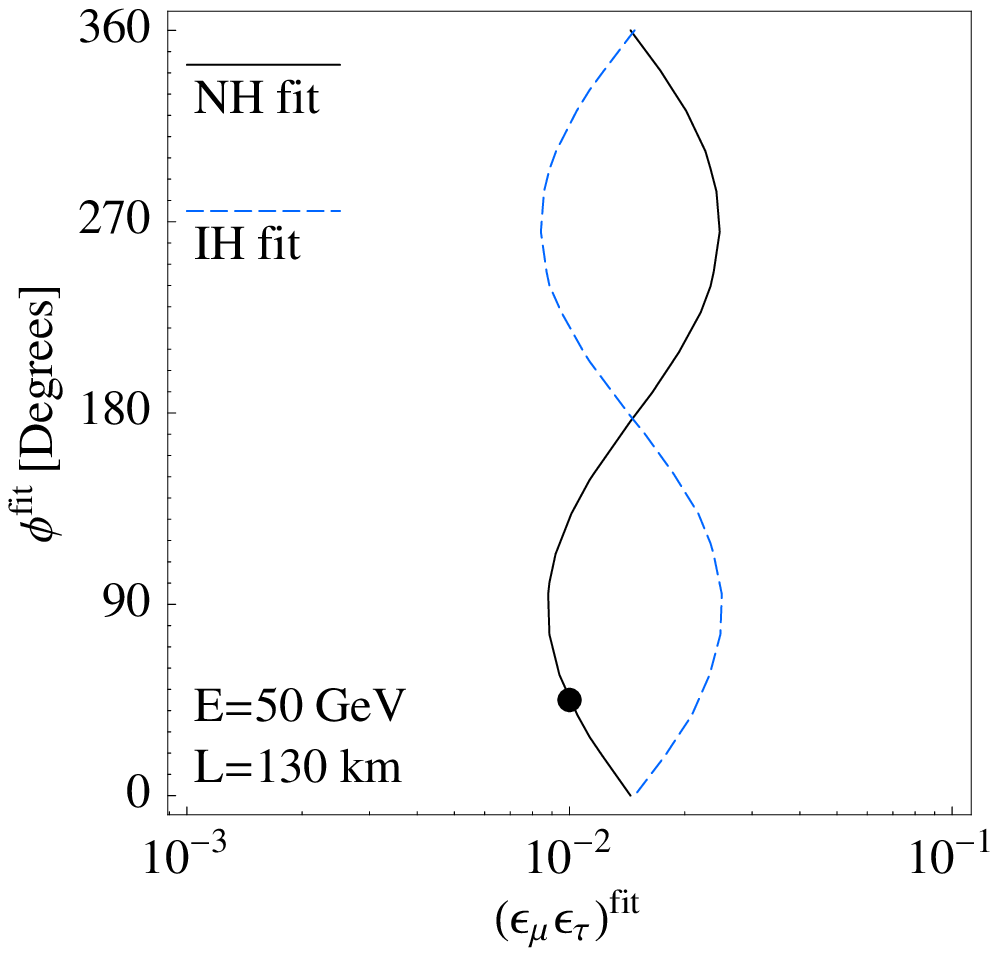}}
\put(12,0){\includegraphics[width=6cm]{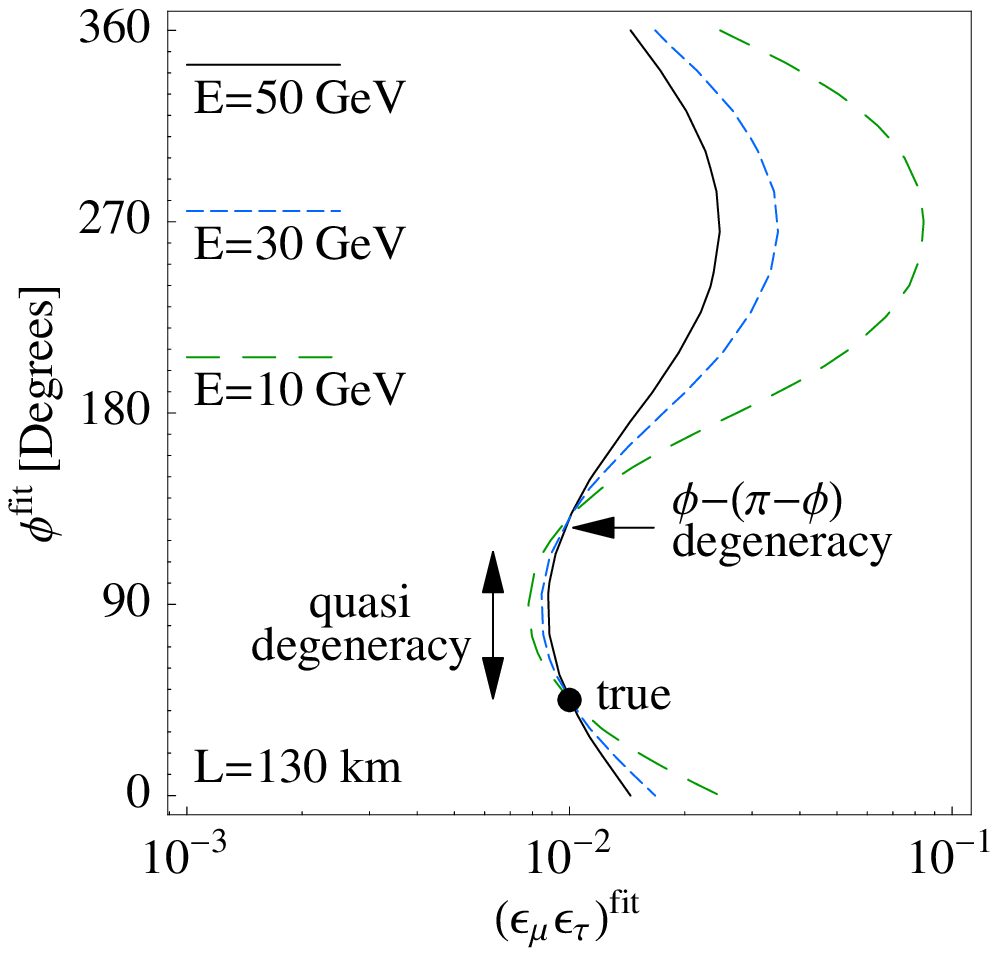}}
\end{picture}
\caption{Equi-probability plots for $\theta_{23}$ degeneracy (left),
for sign of $\Delta m_{31}^{2}$ degeneracy (centre),
and for $\phi$-$(\pi-\phi)$ degeneracy and 
$\epsilon_{\mu} \epsilon_{\tau}$-$\phi$ correlation (right).
The neutrino energy is taken to be 50 GeV and 
the source-detector distance is 130 km.
}
\label{Fig:theta23deg}
\end{figure}

The left panel in Fig.~\ref{Fig:theta23deg}
is for the $\theta_{23}$ degeneracy.
The plot is done for  $E = 50$ GeV, and
the baseline is taken to be 130 km with 2.7 g/cm$^{3}$ as the
matter density although matter effect is not relevant in this setup.
In this plot we draw equi-probability contours in the 
$(\epsmu \epstau)$-$\phi$ plane for two values of 
$\theta_{23}$, 
\begin{align}
\sin^{2}\theta_{23} = \{0.64, ~ 0.36\}.
\end{align}
Here we assume the NH mass spectrum.
The plot shows that the curve with 
$\sin^{2}\theta_{23} = 0.64$ (thin solid)
completely coincides with that of
$\sin^{2}\theta_{23}=0.36$ (thick dashed gray),
and these two cannot be distinguished.
However, this degeneracy does not give rise to any additional 
regions in the 
$(\epsilon_{\mu} \epsilon_{\tau})$-$\phi$ parameter plane
because this degeneracy is not due to the non-unitary parameters, 
i.e., 
the degenerate solutions take the same values of 
$(\epsilon_{\mu} \epsilon_{\tau})$ and $\phi$ 
in the both side of Eq.~\eqref{eq:equiP-condition}, 
\begin{align}
P_{\nu_{\mu} \rightarrow \nu_{\tau}}
       (\theta_{23}, (\epsilon_{\mu} \epsilon_{\tau}), \phi)
       =
       P_{\nu_{\mu} \rightarrow \nu_{\tau}}(\pi/2 - \theta_{23},
      (\epsilon_{\mu} \epsilon_{\tau}), \phi).
\end{align}

The middle panel in Fig.~\ref{Fig:theta23deg} is for the degeneracy
on sign$[\Delta m_{31}^{2}]$-$\phi$.
The solid curve is similar to the curves in
the left panel. 
Here, the standard oscillation parameters are again taken to be
the values in Eq.~\eqref{eq:standardosc-ref} but 
$\sin^{2} \theta_{23}$ is assumed to be 0.5, and 
the NH is adopted in the both side of Eq.~\eqref{eq:equiP-condition}.
In the calculation of the dashed (blue) curve,
the true probability with NH is fitted by 
the probability with the IH mass spectrum.
Therefore,
the condition which is satisfied on the dashed curve is
written as
\begin{align}
 P_{\nu_{\mu} \rightarrow \nu_{\tau}}
 ((\epsilon_{\mu}\epsilon_{\tau})^{\text{fit}}, 
 \phi^{\text{fit}},
 {\rm IH})
 =
 P_{\nu_{\mu} \rightarrow \nu_{\tau}}
 ((\epsilon_{\mu}\epsilon_{\tau})^{\text{true}},
 \phi^{\text{true}},
 \text{NH}).
\label{eq:condition-signdm31Sqdeg}
\end{align}
Although
the dashed curve does not pass through the true value point
which is shown as the black dot in the plot,
the true oscillation probability
can also be reproduced on it.
The shape of the dashed curve is the reflection of the solid
curve at the $\phi=0$ point.

The right panel in Fig.~\ref{Fig:theta23deg}
illustrates the $\phi$-$(\pi-\phi)$ degeneracy and
the $(\epsilon_{\mu} \epsilon_{\tau})$-$\phi$ {\it quasi-}degeneracy.
On each curve, the condition
\begin{align}
 P_{\nu_{\mu} \rightarrow \nu_{\tau}}
 ((\epsilon_{\mu}\epsilon_{\tau})^{\text{fit}}, \phi^{\text{fit}},
 E)
 =
 P_{\nu_{\mu} \rightarrow \nu_{\tau}}
 ((\epsilon_{\mu}\epsilon_{\tau})^{\text{true}},
 \phi^{\text{true}},
 E),
\label{eq:condition-epsilonphideg}
\end{align}
is satisfied,
where the values of the standard oscillation parameters
are again taken as shown in Eq.~\eqref{eq:standardosc-ref}, and
the maximal mixing for $\theta_{23}$ and the NH are assumed in both side
of Eq.~\eqref{eq:condition-epsilonphideg}.
We plot the curves of three cases with the following neutrino energies:
\begin{align}
E = \{ 10,~ 30, ~ 50 \} \qquad \text{[GeV]}.
\end{align}
For a fixed energy all the points on the curve give the same probability 
reflecting the $(\epsmu \epstau)$-$\phi$ degeneracy.  
However, if one considers other illustrative values of 
energies and draws the corresponding equi-probability curves passing 
through the true value point, then in a large region 
of parameter space, the equi-probability curves trace different paths.
Consequently in these regions the  
$(\epsmu \epstau)$-$\phi$ degeneracy can be removed 
by adding the spectral information.  
However, the figure also shows that the 
three curves  cross at two points; one is the true value point
(shown as the black dot),
and the other is the fake solution which is referred as
the $\phi$-$(\pi-\phi)$ degeneracy above for each curve.  
We can also find that at the region between the true
solution and the fake solution,
all three curves take a quite similar path  
indicating in this region the different probabilities for different energies 
have very little dependence on parameters. 
This means that it is hard to resolve
the solutions at this region even with spectral information 
and hence we  mention this as
the {\it quasi}-degeneracy.
We can draw a similar plots as Fig.~\ref{Fig:theta23deg}
with IH as the true hierarchy.

\begin{figure}[tbh]
\unitlength=1cm
\begin{picture}(10,10)
\put(0,0){\includegraphics[width=10cm]{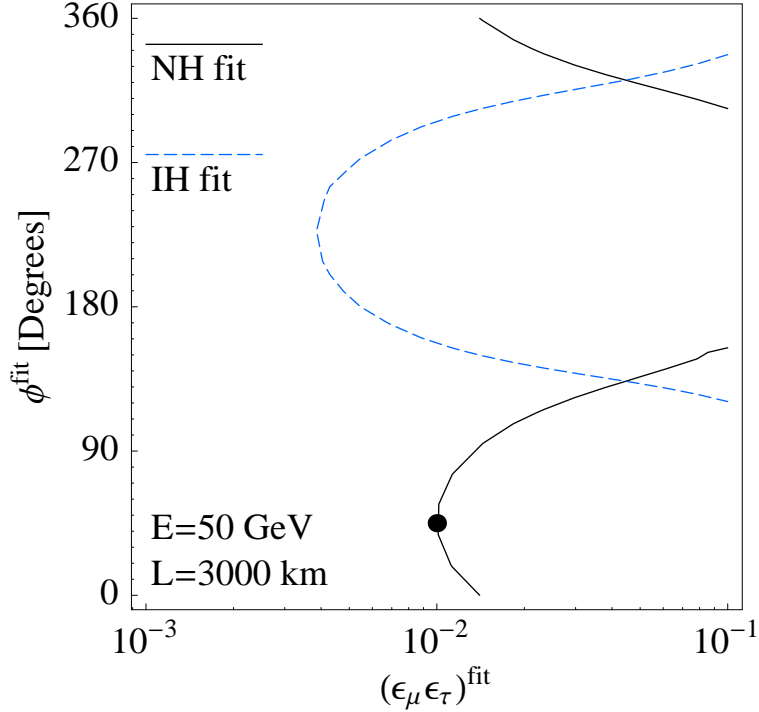}}
\end{picture}
\caption{Equi-probability plot for $E=50$ GeV and $L$ =3000km. 
The solid line denotes the NH fit and the dashed line denotes the IH fit. 
The true value of $(\epsmu \epstau)$ and $\phi$ are marked by the black dot.
The standard oscillation parameters are taken to be 
the values shown in Eq.~\eqref{eq:standardosc-ref} and 
$\theta_{23}$ is assumed to be maximal.}
\label{Fig:prob3000}
\end{figure}

In Fig.~\ref{Fig:prob3000} we plot the equi-probability plot for 
neutrino of energy 50 GeV and $L$ = 3000 km. 
The true hierarchy is assumed to be NH and the true value point 
is again shown as a black dot in the figure. 
The dashed line shows the plot for the IH fit, 
on which the true oscillation probability can be reproduced.
There are two points at which the NH and IH probability crosses each other. 
The conditions for obtaining these points can be worked out from the 
expression Eq.~\eqref{pmutaumat2}.  
In general the condition for degeneracy on these curves can be written as 
\begin{align}
 P_{\nu_{\mu} \rightarrow \nu_{\tau}}
 ((\epsilon_{\mu}\epsilon_{\tau}), \phi, {\rm NH})
 =
 P_{\nu_{\mu} \rightarrow \nu_{\tau}}
 ((\epsilon_{\mu}\epsilon_{\tau})^\prime,
 \phi^\prime, {\rm IH}),
\label{eq:condition-matdeg}
\end{align}
At the point where the NH and IH curves cross the $\epsmu \epstau$ and 
$\phi$ are same for both NH and IH. This gives, 
\be
\tan{\phi} = \frac{a_{\rm NC} L}{2E}.
\label{eq:nheqihinmatter}
\ee 
For $L$ = 3000 km and $E$ = 50 GeV, 
the above gives $\phi \approx 140^o$ and 
$\pi + 140^{o} $ as obtained in the figure.

\section{Numerical Results: 
allowed region on
the $(\epsilon_{\mu} \epsilon_{\tau})$-$\phi$ plane}

In this section we present the results of our numerical analysis.
We first present the allowed regions in the
$(\epsilon_{\mu} \epsilon_{\tau})$-$\phi$ plane
for an OPERA-like detector at a distance of 130 km from a
neutrino factory source and describe
how the degeneracies are realized.
This experimental setup have already been examined in Ref.~\cite{yasuda}.
However, we will pay attention to the degeneracy of the solutions.
Later, we will see that how this degenerate solutions are resolved
including information of matter effect.

\begin{figure}[thb]
\unitlength=1cm
\begin{picture}(18,8)
\put(0,0){\includegraphics[width=8cm]{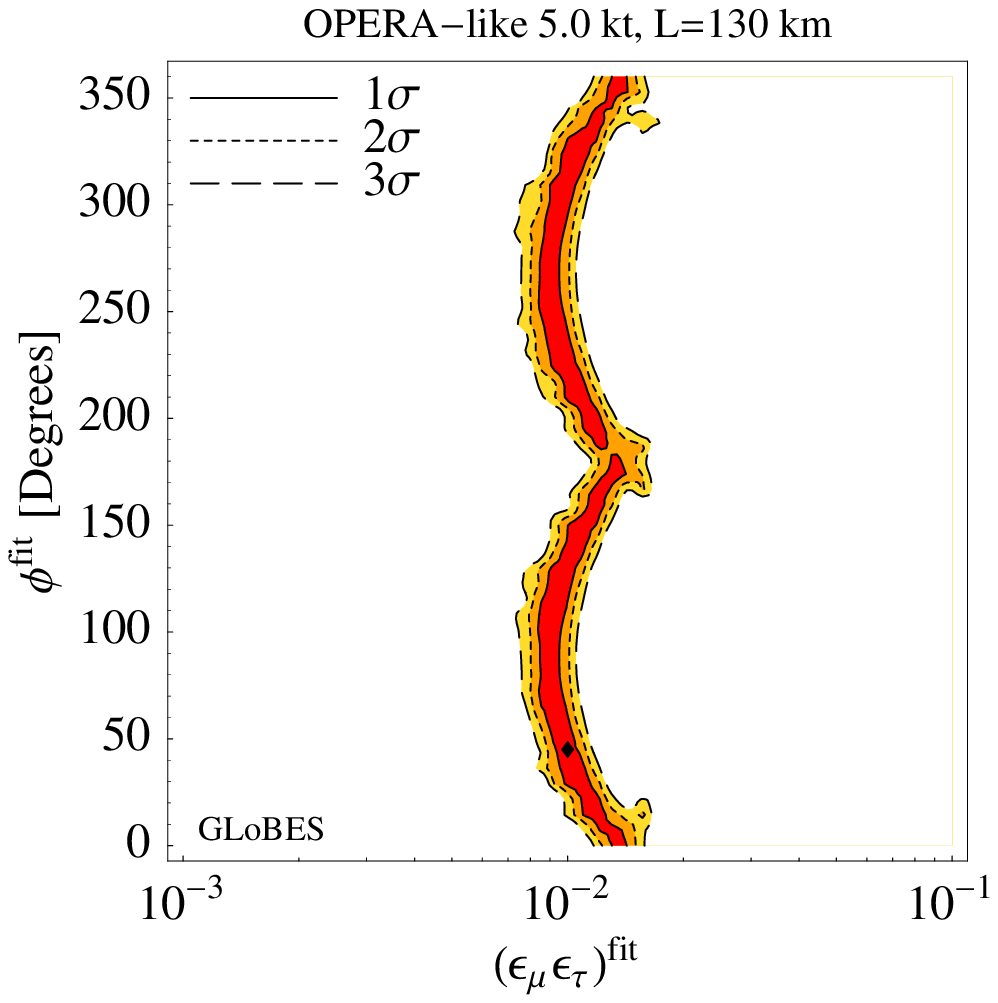}
}
\put(9,0){\includegraphics[width=8cm]{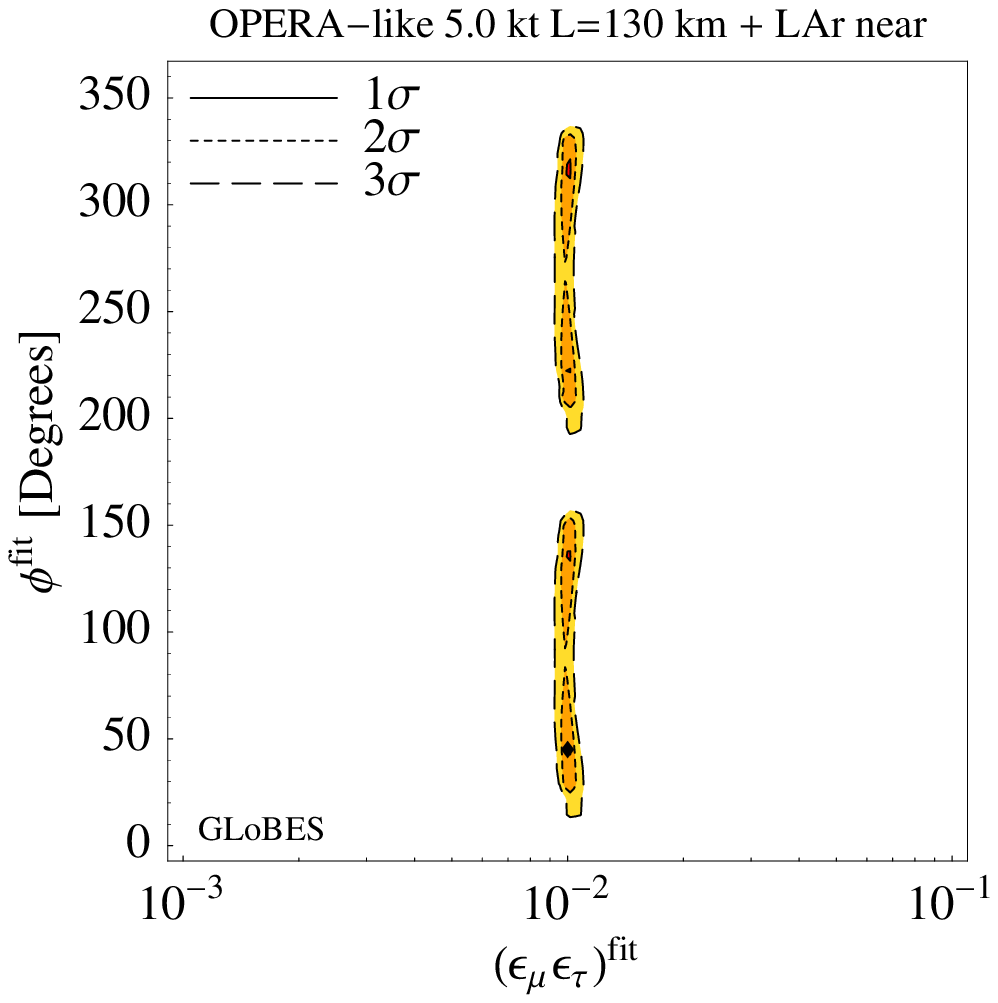}}
\end{picture}
\caption{Allowed regions in the $(\epsilon_{\mu}\epsilon_{\tau})$-$\phi$ plane
for an OPERA-like detector at a distance of 130 km from a Neutrino 
Factory source (left), and for the same setup but with a 0.1 kt Liquid Argon 
near
detector (right). 
}
\label{Fig:OPERA-marginal}
\end{figure}

In Fig.~\ref{Fig:OPERA-marginal},
we plot the $\chi^{2}$ function which is defined as\footnote{%
In the actual implementation, we adopt the Poisson distribution, 
add the appropriately defined priors, and marginalize also over 
the systematic parameters,
following {\sf GLoBES} software \cite{Huber:2004ka,Huber:2007ji}.
}
\begin{align}
\chi^{2}
\left( (\epsilon_{\mu} \epsilon_{\tau})^{\text{fit}}, \phi^{\text{fit}} \right)
=
\min_{\lambda^{\text{fit}}}
\sum_{i}^{\text{bin}}
\left|
 N_{i} (
 \lambda^{\text{true}},
 (\epsilon_{\mu} \epsilon_{\tau})^{\text{true}},
 \phi^{\text{true}})
 -
 N_{i}(\lambda^{\text{fit}},
 (\epsilon_{\mu} \epsilon_{\tau})^{\text{fit}},
 \phi^{\text{fit}})
\right|^{2}
 \bigg/ V_{i},
\end{align}
where $N_{i}$ is the neutrino event number in the $i$-th energy bin,
$\lambda$ represents the standard oscillation parameters and 
$V_{i}$ is the variance which are appropriately defined 
to include the statistical and systematic errors.
Here we adopt the values shown in Eq.~\eqref{eq:standardosc-ref}
for the standard oscillation parameters.
Since it is not 
possible to resolve the $\theta_{23}$ degeneracy
in this experiment
(in the $\nu_{\mu} \rightarrow \nu_{\tau}$ channel),
we take the reference true values for $\theta_{23}$ 
as the maximal. 
The true mass hierarchy is assumed to be NH. 
The parameters for the non-unitary nature are taken as 
shown in Eq.~\eqref{eq:non-uni-ref}. 
The left panel in Fig.~\ref{Fig:OPERA-marginal} 
shows the allowed region in the 
$(\epsilon_{\mu} \epsilon_{\tau})$-$\phi$ plane. 
As discussed earlier, since  the probability in this case is a function of
$\sin^2 2\theta_{23}$, the $\theta_{23}$ octant degeneracy
does not give rise to any additional regions in the
$(\epsilon_{\mu} \epsilon_{\tau})$-$\phi$ plane and the solutions 
for true $\theta_{23}$ and
wrong $\theta_{23}$ occur in the same place.
Here the two solutions (two crescent regions) correspond to 
two choices for the sign of $\Delta m^2_{31}$ in the fit event.
The figure also shows that for each hierarchy there is the 
$\phi$-$(\pi-\phi)$ degeneracy.
The spurious solution corresponding to 
$(\epsilon_{\mu} \epsilon_{\tau})$-$\phi$ degeneracy is 
removed by using the spectrum information. 
There is a weak negative correlation between $\epsmu \epstau$ and $\phi$ 
for each allowed zone. 
We next discuss how one can eliminate the 
degenerate solutions in the 
$(\epsilon_{\mu} \epsilon_{\tau})$-$\phi$ plane 
by combining the experiments at various baselines.  
The remaining degeneracies are 
the sign$[\Delta m_{31}^{2}]$-$\phi$ degeneracy,
the $\phi$-$(\pi-\phi)$ degeneracy, and
the $(\epsilon_{\mu} \epsilon_{\tau})$-$\phi$ quasi-degeneracy. 
The right plot of Fig.~\ref{Fig:OPERA-marginal}
shows the combining results of an OPERA-like detector 
at 130 km baseline and a 0.1 kton liquid Argon (LAr) type near 
detector which is located at $L=2$ km\footnote{%
A 0.1 kt LAr detector as a near detector 
has been discussed in Ref.~\cite{ic}.}.
The probability at the near detector depends on $\epsmu \epstau$ only. 
Thus combining with this experiment helps to narrow down the allowed 
region but the degeneracies still exist. The correlation between $\phi$ and 
$\epsmu \epstau$ is now almost vanishing.

\begin{figure}[thb]
\unitlength=1cm
\begin{picture}(18,8)
\put(0,0){\includegraphics[width=8cm]{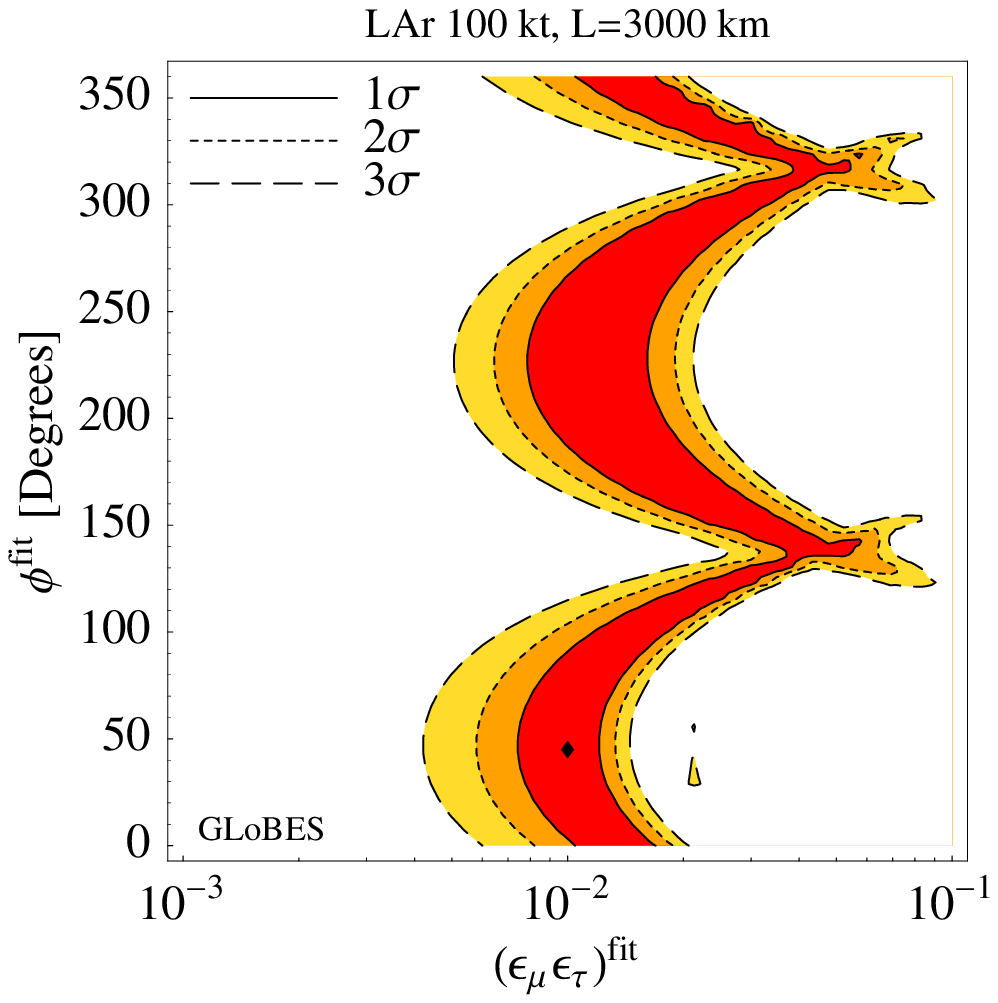}}
\put(9,0){\includegraphics[width=8cm]{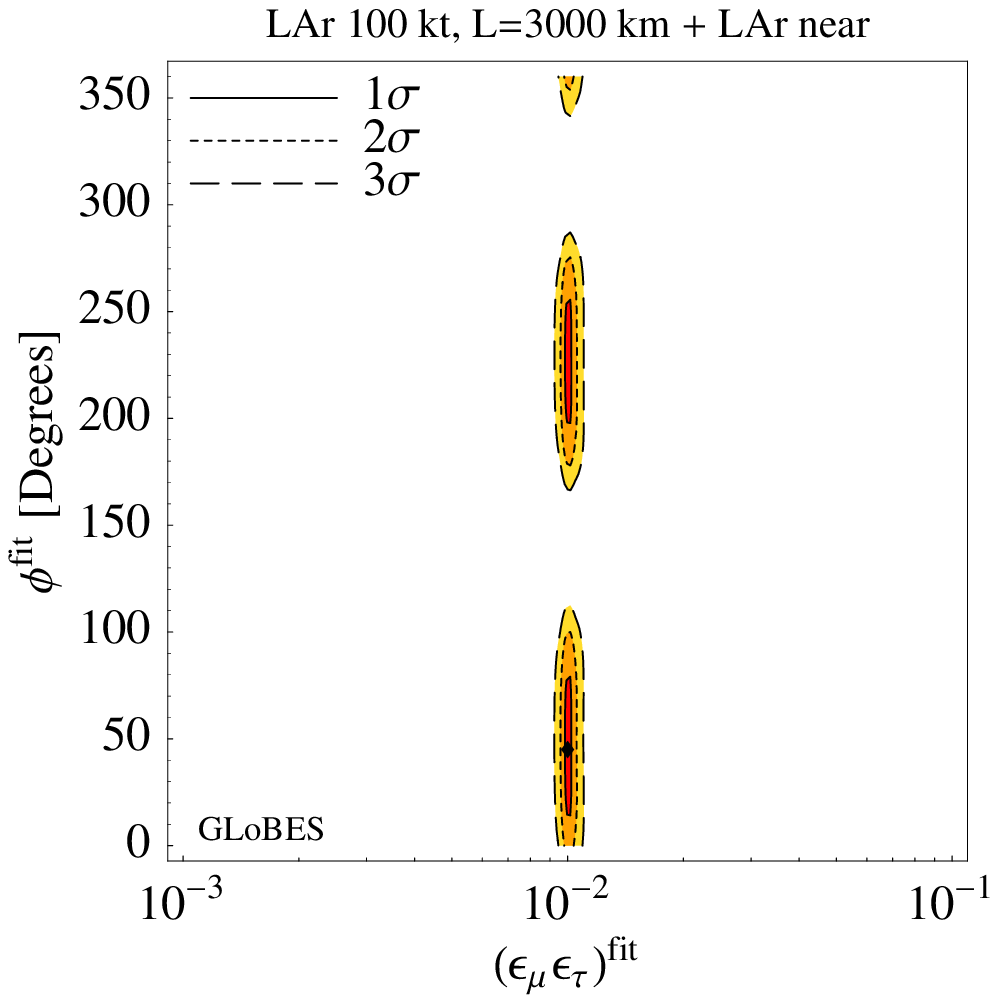}}
\end{picture}
\caption{Same as Fig.~\ref{Fig:OPERA-marginal}
but we assume a 100 kt LAr detector at $L=3000$ km in the left panel.
In the right panel, the same near detector as
 Fig.~\ref{Fig:OPERA-marginal} is added.
}
\label{Fig:3000_nh+ih}
\end{figure}

In left panel of Fig.~\ref{Fig:3000_nh+ih} we plot the allowed regions for 
the combination of a neutrino factory and a 100 kton LAr type far detector 
which is located at $L=3000$ km.
A comparison of this figure with the equi-probability plot 
Fig.~\ref{Fig:prob3000} reveals that the middle region 
in Fig.~\ref{Fig:3000_nh+ih} corresponds to the IH fit. 
As discussed in the previous section 
the main contribution of matter effect depends 
on $\cos \phi$ which is different from the case in vacuum.
Therefore, the $\phi$-$(\pi - \phi)$   
degeneracy as well as sign$[\Delta m^2_{31}]$-$\phi$ degeneracy 
can be removed by this matter term. This is reflected in the figure.
However the probabilities for NH and IH can still be equal when the 
condition Eq.~\eqref{eq:condition-matdeg} is satisfied. 
This gives rise to the middle region in Fig.~\ref{Fig:3000_nh+ih}.
There is a positive correlation in this case between $\epsmu \epstau$ 
and $\phi$ for each allowed region. 
The result combining the near detector
is shown as the right panel in Fig.~\ref{Fig:3000_nh+ih}.
This helps to reduce the uncertainty in the $\epsmu \epstau$
and since $\phi$ is a variable correlated with 
$\epsmu \epstau$, the uncertainty on $\phi$ is also reduced.
Therefore, the allowed regions for each hierarchy becomes much narrower 
as compared to the left panel.
The allowed regions now are almost parallel to the $\phi$ axis.  
The inclusion of the 3000 km removes the 
the $\phi$-$(\pi-\phi)$ degeneracy 
of Fig.~\ref{Fig:OPERA-marginal} for each
hierarchy. 
In addition, in such a long baseline experiment, 
it would be possible to obtain information on the
sign$[\Delta m_{31}^{2}]$ from the other channels like $\nu_{e}
\rightarrow \nu_{\mu}$.
Including it, we could remove the wrong hierarchy solution 
and solve the all degeneracies.

\section{Conclusions} 
Non-unitary mixing matrix is a generic feature for theories with 
mixing between neutrinos and heavy states and provides a window 
to probe physics at high scale. 
In this paper we have studied the possibility of probing non-unitarity 
of neutrino mixing matrix at neutrino factories. 
We considered the $\nu_{\mu} \rightarrow \nu_{\tau}$ channel 
and detectors at a distance of 2 km, 130 km and 3000 km from the source. 
We show that for the $\nu_{\mu} \rightarrow \nu_{\tau}$ channel at 130 km,
there can be degenerate solutions even for $\theta_{13}$=0 in the 
$(\epsilon_{\mu} \epsilon_{\tau})$-$\phi$ plane 
where $\phi$ and $\epsmu \epstau$ are the 
phase and moduli of the unitarity violation parameter. 
The degenerate solutions in the 
$(\epsilon_{\mu} \epsilon_{\tau})$-$\phi$ plane are due to 
\begin{itemize}
\item
$(\Delta m^2_{31} > 0, \phi) 
\rightarrow (\Delta m^2_{31} < 0, - \phi)$

\item
$\phi \rightarrow \pi -\phi$ 

\item
$(\epsmu \epstau, \phi) \rightarrow 
     ((\epsmu\epstau)^\prime, \phi^{\prime})$ 

\end{itemize} 
For a detector at distance 130 km from a neutrino factory source 
the last degeneracy can be removed using spectral information 
and no additional disconnected solution appear.
By adding an experiment at 2 km the correlation between $\phi$ and 
$\epsmu \epstau$ can be reduced and the allowed ranges narrow down. 
For the 3000 km experiment the matter effects are relevant and this
removes the first and second degeneracy listed above.  
However although the hierarchy degeneracy listed above gets removed, 
there can still be the degeneracy where probabilities 
for NH and IH give same values. 
If we consider only the 3000 km experiment then there is a greater
correlation between $\epsmu \epstau$-$\phi$ and the allowed regions
are larger as compared to the 130 km experiment. 
However, addition of the 2 km experiment to this 
reduces this correlation and the allowed regions become narrower. 
Although we have concentrated on 
the $\nu_{\mu} \rightarrow \nu_{\tau}$ channel in this study,
if we combine the other channel like $\nu_{e} \rightarrow \nu_{\mu}$, 
we can obtain information on 
the hierarchy and then the allowed regions further reduce in size.
 
\begin{acknowledgments}
We would like to thank A. Bandyopadhyay, P. Ghoshal, 
J. Kopp, W. Rodejohann, R. Singh and S. Umashankar 
for useful discussions  
and gratefully acknowledge 
M. Lindner and the particle and astroparticle physics group 
in the Max-Planck-Institut f\"ur Kernphysik for hospitality. 
S.G. acknowledges the Alexander-von-Humboldt Foundation for support.
\end{acknowledgments}

\appendix

\section{Analytic Formulae} 
In this section, we derive the expression of 
$\nu_\mu \rightarrow \nu_\tau$ oscillation
probability in matter of constant density under 
some simplifying assumptions.
If the neutrino mixing matrix is non-unitary then,
although we can get the canonical form of the kinetic energy in the mass basis
in terms of the flavour states the kinetic term is not diagonal.
Therefore it is more appropriate here to consider the
neutrino propagation equation in the mass basis.
The neutrino propagation Hamiltonian in matter
can be generally represented in the vacuum mass eigenbasis
as follows,
\begin{align}
H_{ij}
 =&
 \frac{1}{2E}
 \left\{
 \begin{pmatrix}
  0 & & \\
  & \Delta m_{21}^{2} & \\
  & & \Delta m_{31}^{2}
 \end{pmatrix}
 +
 a_{\rm CC}
 \begin{pmatrix}
 {|{W_{e 1}}|}^{2} &
    {W^{*}_{e1}}
 {W_{e 2}} &
    {W^{*}_{e 1}}
 {W_{e 2}} \\
  {W^{*}_{e 2}}
 {W_{1e}}
  &
  {|{W_{e2 }}|^2}
  &
    {W^{*}_{e2}}
 {W_{e 3}} \\
    {W^{*}_{e3}}
  {W_{1e}}
  &
  {W^{*}_{e3}}
  {W_{e2}}
  &
  {|{W_{e 3}}|}^{2}
 \end{pmatrix} \right. \nonumber \\
 &
 \left.
 \hspace{2cm}
+ a_{\rm NC}
 \begin{pmatrix}
  \displaystyle
  \sum_{\gamma=e,\mu,\tau}
  {|{W_{\gamma 1}}|}^{2}
  &
   \displaystyle
  \sum_{\gamma=e,\mu,\tau}
  {W^{*}_{ \gamma 1}}
 {W_{\gamma 2}}
  &
  \displaystyle
  \sum_{\gamma=e,\mu,\tau}
 {W^{*}_{ \gamma 1}}
 {W_{\gamma 3}}
 \\
  \displaystyle
 \sum_{\gamma=e,\mu,\tau}
 {W^{*}_{\gamma 2}}
 {W_{\gamma 1}}
  &
  \displaystyle
   \sum_{\gamma=e,\mu,\tau}
 {|{W_{\gamma}}^{2}|}^{2}
  &
  \displaystyle
   \sum_{\gamma=e,\mu,\tau}
 {W^{*}_{\gamma 2}}
 {W_{\gamma 3 }}
  \\
  \displaystyle
   \sum_{\gamma=e,\mu,\tau}
 {W^{*}_{\gamma 3}}
  {W_{\gamma 1}}
  &
  \displaystyle
   \sum_{\gamma=e,\mu,\tau}
 {W^{*}_{\gamma 3}}
 {W_{2 \gamma}}
  &
   \displaystyle
   \sum_{\gamma=e,\mu,\tau}
  {|{W_{\gamma  3}}|}^{2}
 \end{pmatrix}
 \right\},
\label{eq:H-in-matter}
\end{align}
where
$a_{\rm CC} \equiv 2\sqrt{2} E G_{F} n_{e} $,
$a_{\rm NC} \equiv -\sqrt{2} E G_{F} n_{N} = - a_{\rm CC}/2$
are the charged and neutral current potentials respectively.
We obtain simplified analytic expressions for the probability by solving
the above equations in the limit
$\theta_{13} \rightarrow 0$, $\Delta m_{21}^{2}/\Delta m_{31}^{2} 
\rightarrow 0$. 
In this limit the propagation Hamiltonian 
in the vacuum mass eigenbasis is
\begin{align} 
H_{ij} = (H_{0})_{ij} 
 + (H_{\epsilon_{\mu} \epsilon_{\tau}})_{ij} 
 + (H_{\epsilon_\mu^2})_{ij} 
 + (H_{\epsilon_\tau^2})_{ij}
\end{align}
where 
\begin{align}
{(H_{0})_{ij}}
=&
 \frac{1}{2E}
 \left\{
  \begin{pmatrix}
  0 & & \\
  & 0 & \\
  && \Delta m_{31}^{2}
 \end{pmatrix}
 +
 a_{\rm CC}
 \begin{pmatrix}
  c_{12}^{2} & c_{12} s_{12} & 0 \\
  c_{12} s_{12} & s_{12}^{2} & 0 \\
  0 & 0 & 0
 \end{pmatrix}
\right\}, \\
 {(H_{\epsilon_{\mu} \epsilon_{\tau}})_{ij}}
 =&
 -\frac{a_{\rm NC}}{2E}
 \epsilon_\mu \epsilon_\tau
 \begin{pmatrix}
  2 c_{23} s_{23} s_{12}^{2} c_{\phi}
  &
  - 2c_{23} s_{23} c_{12} s_{12} c_{\phi}
  &
  s_{12} (c_{23}^{2} {\rm e}^{-{\rm i} \phi} - s_{23}^{2}
  {\rm e}^{{\rm i} \phi}) \\
  - 2c_{23} s_{23} c_{12} s_{12} c_{\phi}
  &
  2 c_{23} s_{23} c_{12}^{2} c_{\phi}
  &
  -c_{12} (c_{23}^{2} {\rm e}^{-{\rm i} \phi} - s_{23}^{2}
  {\rm e}^{{\rm i} \phi}) \\
  s_{12} (c_{23}^{2} {\rm e}^{{\rm i} \phi} - s_{23}^{2}
  {\rm e}^{-{\rm i} \phi})
 &
  -c_{12} (c_{23}^{2} {\rm e}^{{\rm i} \phi} - s_{23}^{2}
  {\rm e}^{-{\rm i} \phi})
  &
  -
  2 c_{23} s_{23}
  c_{\phi}
 \end{pmatrix}, \\
 (H_{\epsilon_{\mu}^{2}})_{ij}
 =& -\frac{a_{\rm NC}}{2E}
  \epsilon_{\mu}^{2}
 \begin{pmatrix}
   s_{12}^{2} c_{23}^{2}
  &
  -s_{12} c_{12} c_{23}^{2}
  &
  -s_{12} s_{23} c_{23} \\
  - s_{12} c_{12} c_{23}^{2}
  &
  c_{12}^{2} c_{23}^{2}
  &
 c_{12} s_{23} c_{23} \\
  -s_{12} s_{23} c_{23}
  &
  c_{12} s_{23} c_{23}
  &
  s_{23}^{2}
 \end{pmatrix}, \\
 (H_{\epsilon_{\tau}^{2}})_{ij}
 =&
 -\frac{a_{\rm NC}}{2E}
 \epsilon_{\tau}^{2}
 \begin{pmatrix}
    s_{12}^{2} s_{23}^{2}
  &
  -s_{12} c_{12} s_{23}^{2}
  &
  s_{12} s_{23} c_{23} \\
  - s_{12} c_{12} s_{23}^{2}
  &
  c_{12}^{2} s_{23}^{2}
 &
  -c_{12} s_{23} c_{23} \\
  s_{12} s_{23} c_{23}
  &
  -c_{12} s_{23} c_{23}
  &
  c_{23}^{2}
 \end{pmatrix},
\end{align}
up to the second order of the epsilon parameters.
In writing the above a part proportional to  unit matrix 
is omitted as it contributes to overall phase.
The Hamiltonian is separated  into two parts ---
the zeroth order part $H_{0}$ which includes $\Delta m_{31}^{2}$
and $a_{\rm CC}$,
and perturbations
$H_{\epsilon_{\mu} \epsilon_{\tau}}$,
$H_{\epsilon_{\mu}^{2}}$, and $H_{\epsilon_{\tau}^{2}}$,
induced by the non-unitarity.
Note that
the non-unitarity effects appear
always at the second order (or higher than that)
of the $\epsilon_{\alpha}$ parameters.

Treating $H_{\epsilon_{\mu} \epsilon_{\tau}}$, $H_{\epsilon_{\mu}^{2}}$
and $H_{\epsilon_{\tau}^{2}}$ as perturbations,
the amplitude of the neutrino oscillation
from a vacuum mass eigenstate $\nu_{i}$ to the other vacuum
mass eigenstate $\nu_{j}$ can be written as
\begin{align}
{S_{ji}} =
 {(S_{0})_{ji}}
 +
 {(S_{\epsilon_{\mu} \epsilon_{\tau}})_{ji}}
 +
 {(S_{\epsilon_{\mu}^{2}})_{ji}},
 +
 {(S_{\epsilon_{\tau}^{2}})_{ji}},
\end{align}
where
$S_{0}$ is the zeroth order part, and
$S_{\epsilon_{\mu} \epsilon_{\tau}}$, 
$S_{\epsilon_{\mu}^{2}}$, 
and $S_{\epsilon_{\tau}^{2}}$
correspond to the
amplitudes with perturbations of
$H_{\epsilon_{\mu} \epsilon_{\tau}}$, 
$H_{\epsilon_{\mu}^{2}}$, and $H_{\epsilon_{\tau}^{2}}$
respectively,
which are calculated to be
\begin{align}
 {(S_{0})_{ji}}
 =& {({\rm e}^{-{\rm i} H_{0} L})_{ji}}, \\
 {(S_{\epsilon_{\mu} \epsilon_{\tau}})_{ji}}
 =&
 {({\rm e}^{-{\rm i} H_{0} L})_{jk}}
 (-{\rm i})
 \int_{0}^{L} {\rm d} x
 {({\rm e}^{+{\rm i} H_{0} x})_{k l}}
 {(H_{\epsilon_{\mu} \epsilon_{\tau}})_{l m}}
 {({\rm e}^{-{\rm i} H_{0} x})_{m i}},
 \label{eq:S-eta-define}\\
 {(S_{\epsilon_{\mu}^{2}})_{j i}}
 =&
 {({\rm e}^{-{\rm i} H_{0} L})_{j k}}
 (-{\rm i})
 \int_{0}^{L} {\rm d} x
 {({\rm e}^{+{\rm i} H_{0} x})_{k l}}
 {(H_{\epsilon_{\mu}^{2}})_{l m}}
 {({\rm e}^{-{\rm i} H_{0} x})_{m i}}, \\
 {(S_{\epsilon_{\tau}^{2}})_{j i}}
 =&
 {({\rm e}^{-{\rm i} H_{0} L})_{j k}}
 (-{\rm i})
 \int_{0}^{L} {\rm d} x
 {({\rm e}^{+{\rm i} H_{0} x})_{k l}}
 {(H_{\epsilon_{\tau}^{2}})_{l m}}
 {({\rm e}^{-{\rm i} H_{0} x})_{m i}},
\end{align}
Note that these amplitudes describe
the transition between two vacuum mass eigenstates,
$\nu_{i}$ and $\nu_{j}$ and
a transition between flavour states can be obtained
by sandwiching them by the flavour states\footnote{
We underscore that, strictly speaking, 
this method  has to be followed
as it is not correct to write the the neutrino propagation 
in the flavour basis 
because the flavour states do not form a complete set for the
propagation Hamiltonian.} 
which are
described as \cite{ll,Holeczek:2007kk,Giunti}
\begin{align}
 |\nu_{\alpha} \rangle
 =
 \frac{1}{\displaystyle 
 \sqrt{\sum_{j=1}^{\text{light}} {|W_{\alpha j}|}^{2} } }
 \sum_{i=1}^{\text{light}}
 W_{\alpha i}^{*}|\nu_{i} \rangle
\end{align} 
The oscillation probability between two flavour states
$\nu_{\alpha}$ and $\nu_{\beta}$
is derived as
\begin{align}
 P_{\nu_{\alpha} \rightarrow \nu_{\beta}}
 =&
 \left|
 \frac{1}{\displaystyle 
 \sqrt{\sum_{l=1}^{\text{light}} {|W_{\beta l}|}^{2} } }
 {W_{\beta j}}
 {\left(
 S_{0} + S_{\epsilon_{\mu} \epsilon_{\tau}} + S_{\epsilon_{\mu}^{2}} + S_{\epsilon_{\tau}^{2}}
 \right)_{ji}}
 \frac{1}{\displaystyle 
 \sqrt{\sum_{k=1}^{\text{light}} {|W_{\alpha k}|}^{2} } }
 (W^{\dagger})_{i \alpha}
 \right|^{2} \nonumber \\
 =&
 \frac{1}{N_{\alpha} N_{\beta}}
 \left[
 {\left| {(S_{0})_{\beta \alpha}} \right|}^{2}
 +
 2 {\rm Re}
 [
 {(S_{0}^{*})_{\beta \alpha}}
 {(S_{\epsilon_{\mu} \epsilon_{\tau}})_{\beta \alpha }}
 ]
 +
 2 {\rm Re}
 [
 {(S_{0}^{*})_{\beta \alpha}}
 {(S_{\epsilon_{\mu}^{2}})_{\beta \alpha}}
 ]
 +
 2 {\rm Re}
 [
 {(S_{0}^{*})_{\beta \alpha }}
 {(S_{\epsilon_{\tau}^{2}})_{\beta \alpha }}
 ]
 \right] + \mathcal{O}(\epsilon^{4}),
\end{align}
up to the first order perturbations. 
In the following,
we will calculate each oscillation amplitude.

Diagonalizing the zeroth order Hamiltonian $H_{0}$,
we obtain the mass squared eigenvalues and
the mixing matrix ${(V_{0})_{i\tilde{j}}}$
which connects the vacuum mass eigenbasis $\nu_{i}$ with
the mass eigenbasis in matter $\nu_{\tilde{j}}$, and in the limit
which we adopt here, they take  the following simple forms
\begin{align}
 (H_{0})_{\tilde{k}}
 = \text{diag}(a_{\rm CC}, 0, \Delta m_{31}^{2})
 =
 {(V_{0}^{\dagger})_{\tilde{k} j}}
 {(H_{0})_{j i}}
 {(V_{0})_{i \tilde{k}}},
\end{align}
where
\begin{align}
 (V_{0})_{i \tilde{j}} =
 \begin{pmatrix}
  c_{12} & -s_{12} &  \\
  s_{12} & c_{12} & \\
  && 1
 \end{pmatrix}.
\end{align}
Therefore,
the zeroth order amplitude in the vacuum mass eigenbasis
becomes
\begin{align}
 {(S_{0})_{j i}}
 =
 {(V_{0})_{j \tilde{k}}}
 \begin{pmatrix}
  {\rm e}^{- {\rm i} \frac{a_{\rm CC}L}{2E} } &&\\
  & 1 & \\
  &&  {\rm e}^{- {\rm i} \frac{\Delta m_{31}^{2}L}{2E}}
 \end{pmatrix}
 {(V_{0}^{\dagger})_{\tilde{k} i}},
\end{align}
and that for the transition between two flavour states
is
\begin{align}
 {(S_{0})_{\beta \alpha}}
 =
 W_{\beta j} (S_{0})_{j i} (W^{\dagger})_{i \alpha}.
\end{align}
The oscillation probability at the zeroth order becomes
\begin{align}
 P^{\text{0th}}_{\nu_{\mu} \rightarrow \nu_{\tau}}
 =
 \sin 2\theta_{23}
 \left(
 \sin 2\theta_{23}
 +
 2 \epsilon_{\mu} \epsilon_{\tau} \cos 2 \theta_{23} \cos\phi
 \right)
 \sin^{2} \frac{\Delta m_{31}^{2} L}{4E}
 +
 \epsilon_{\mu} \epsilon_{\tau}
 \sin \phi
 \sin 2\theta_{23}
 \sin \frac{\Delta m_{31}^{2} L }{2E} + \mathcal{O}(\epsilon^{3}),
 \label{eq:Pmutau-0th}
\end{align}
which is the same as the formula in the vacuum case.

Next, let us turn to the perturbation terms.
First one is the amplitude of $S_{\epsilon_{\mu} \epsilon_{\tau}}$.
According to Eq.~\eqref{eq:S-eta-define}, we can calculate it as
\begin{align}
{(S_{\epsilon_{\mu} \epsilon_{\tau}})_{j i}}
 =
 \epsilon_{\mu} \epsilon_{\tau}
 \left(
 {\rm i}
 \frac{a_{\rm NC}}{2E}
 \right)
 {(V_{0})_{j \tilde{l}}}
 \begin{pmatrix}
  0 & 0 & 0 \\
  0
  &
  - s_{2\times 23} c_{\phi} L
  &
  \mathcal{A}
  \frac{2E}{{\rm i} \Delta m_{31}^{2}}
  \left(
  1-{\rm e}^{-{\rm i} \frac{\Delta m_{31}^{2} L}{2E}}
  \right) \\
  0
  &
\mathcal{A}^{*}
  \frac{2E}{{\rm i} \Delta m_{31}^{2}}
  \left(
  1-{\rm e}^{-{\rm i} \frac{\Delta m_{31}^{2} L}{2E}}
  \right)
  &
  s_{2\times 23} c_{\phi} L {\rm e}^{-{\rm i} \frac{\Delta m_{31}^{2} L }{2E}}
 \end{pmatrix}
 {(V_{0}^{\dagger})_{\tilde{k} i}},
\end{align}
where the parameters $\mathcal{A}$ is defined as
\begin{align}
 \mathcal{A}
 \equiv
 (c_{23}^{2} {\rm e}^{-{\rm i}\phi} - s_{23}^{2} {\rm e}^{{\rm i}\phi}),
 \end{align}
and $s_{2\times 23} \equiv \sin 2 \theta_{23}$.
The amplitude for $\nu_{\mu} \rightarrow \nu_{\tau}$ transition is
reduced to
\begin{align}
 {(S_{\epsilon_{\mu} \epsilon_{\tau}})_{\tau \mu}}
 =&
 \epsilon_{\mu} \epsilon_{\tau}
 \left[
  {\rm i}
  \frac{a_{\rm NC} L}{4E}
  s_{2\times 23}^{2} c_{\phi}
 \left(1+ {\rm e}^{-{\rm i} \frac{\Delta m_{31}^{2} L}{2E}} \right)
 +
 \frac{a_{\rm NC}}{\Delta m_{31}^{2}}
 \left(
 {\rm e}^{{\rm i} \phi} - s_{2\times 23}^{2} c_{\phi}
 \right)
 \left(1- {\rm e}^{-{\rm i} \frac{\Delta m_{31}^{2} L}{2E}} \right)
 \right] + \mathcal{O}(\epsilon^{3}).
\end{align}
The contribution to the oscillation probability is calculated to be
\begin{align}
 2 {\rm Re}[{(S_{0}^{*})_{\tau \mu}} 
 {(S_{\epsilon_{\mu} \epsilon_{\tau}})_{\tau \mu}}]
 =&
 -
 \epsilon_{\mu}
 \epsilon_{\tau} 
 \left( \frac{a_{\rm NC} L}{2E} \right)
 s_{2 \times 23}^{3} c_{\phi}
 \sin \frac{\Delta m_{31}^{2} L}{2E}
 -
 4 
 \epsilon_{\mu}
 \epsilon_{\tau}
 \left(\frac{a_{\rm NC}}{\Delta m_{31}^{2}}\right)
 s_{2\times 23} c_{2 \times 23}^{2} c_{\phi}
 \sin^{2} \frac{\Delta m_{31}^{2} L}{4E},
\label{eq:dP-eta}
\end{align}
up to the second order of the $\epsilon$ parameters.
The contributions from $S_{\epsilon_{\mu}^{2}}$ and
 $S_{\epsilon_{\tau}^{2}}$ can also be calculated with the same way,
which are
\begin{align}
 2 {\rm Re}[{(S_{0}^{*})_{\tau \mu}} {(S_{\epsilon_{\mu}^{2}})_{\tau \mu}}]
 +
 2 {\rm Re}[{(S_{0}^{*})_{\tau \mu}}
 {(S_{\epsilon_{\tau}^{2}})_{\tau \mu}}]
 =&
 \left(
 \frac{a_{\rm NC} L}{4E}
 \right)
 s_{2\times 23}^{2} c_{2\times 23}
 (\epsilon_{\mu}^{2} - \epsilon_{\tau}^{2})
 \sin \frac{\Delta m_{31}^{2} L}{2E} \nonumber \\
 &-2
 \left(\frac{a_{\rm NC}}{\Delta m_{31}^{2}} \right)
 s_{2\times23}^{2} c_{2\times 23}
 (\epsilon_{\mu}^{2} -\epsilon_{\tau}^{2})
 \sin^{2} \frac{\Delta m_{31}^{2} L}{4E}. 
\label{eq:dP-epsilonSq}
\end{align}

From Eqs.~\eqref{eq:Pmutau-0th}, \eqref{eq:dP-eta}
and \eqref{eq:dP-epsilonSq},
the oscillation probability for $\nu_{\mu} \rightarrow
\nu_{\tau}$
in matter can be expressed as
\begin{align}
 P_{\nu_{\mu} \rightarrow \nu_{\tau}}
 =&
 \sin 2\theta_{23}
 \left(
 \sin 2 \theta_{23}
 +
 2 \epsilon_{\mu} \epsilon_{\tau} \cos 2\theta_{23} \cos\phi
 \right)
 \sin^{2} \frac{\Delta m_{31}^{2} L}{4E}
 +
 \epsilon_{\mu} \epsilon_{\tau}
 \sin \phi
 \sin 2\theta_{23}
\sin \frac{\Delta m_{31}^{2} L }{2E} \nonumber \\
&-
 \epsilon_{\mu} \epsilon_{\tau}
 \left( \frac{a_{\rm NC} L}{2E} \right)
 \sin^{3} 2 \theta_{23} \cos\phi
 \sin \frac{\Delta m_{31}^{2} L}{2E} 
 -
 4 \epsilon_{\mu} \epsilon_{\tau}
 \left(\frac{a_{\rm NC}}{\Delta m_{31}^{2}}\right)
 \sin 2 \theta_{23} \cos^{2} 2 \theta_{23} \cos\phi
 \sin^{2} \frac{\Delta m_{31}^{2} L}{4E}
 \nonumber \\
 &
 -2
 \left(\frac{a_{\rm NC}}{\Delta m_{31}^{2}} \right)
 \sin^{2} 2\theta_{23} \cos 2 \theta_{23}
 (\epsilon_{\mu}^{2} -\epsilon_{\tau}^{2})
 \sin^{2} \frac{\Delta m_{31}^{2} L}{4E} 
 +
 \left(
 \frac{a_{\rm NC} L}{4E}
 \right)
 \sin^{2} \theta_{23} \cos 2 \theta_{23}
 (\epsilon_{\mu}^{2} - \epsilon_{\tau}^{2})
 \sin \frac{\Delta m_{31}^{2} L}{2E} \nonumber \\
 &+
 \mathcal{O}(\epsilon^{3})
 +
 \mathcal{O}(s_{13})
 +
 \mathcal{O}(\Delta m_{21}^{2}/\Delta m_{31}^{2}).
\end{align}
Since $\theta_{23} \simeq \pi/4$, we can omit the terms which
are proportional to $\cos 2\theta_{23}$, and
finally, it reduces to
\begin{align}
 P_{\nu_{\mu} \rightarrow \nu_{\tau}}
  =&
 \sin^{2} 2\theta_{23}
 \sin^{2} \frac{\Delta m_{31}^{2} L}{4E}
 +
 \epsilon_{\mu} \epsilon_{\tau}
 \sin 2\theta_{23}
 \sin \phi
 \sin \frac{\Delta m_{31}^{2} L }{2E} 
 -
 \epsilon_{\mu} \epsilon_{\tau}
 \left( \frac{a_{\rm NC} L}{2E} \right)
 \sin^{3} 2 \theta_{23}
 \cos\phi
 \sin \frac{\Delta m_{31}^{2} L}{2E}.
\end{align}

\section{Experimental setups in Numerical Calculations} 

The numerical work is performed using  
{\sf GLoBES} software~\cite{Huber:2004ka,Huber:2007ji}
which is modified for our purpose.
We consider a neutrino factory as the source for $\nu_\mu$s  
based on {\sf NuFact2} from Ref.~\cite{Huber:2002mx}.
The number of the decay muon is assumed to be $1.06 \cdot 10^{21}$ per
year and four years of running is being considered. 
Here, we concentrate on one polarity of the muon ($\mu^{-}$).
The stored muon is accelerated to 50 GeV.

We perform a binned $\chi^2$-analysis with
energy window  from 1 to 50 GeV and width of each bin as 1 GeV.   
The signal event rate in the $i$-th energy bin is calculated as
\begin{align}
N_{i}^{\text{signal}}
=
 \int_{E_{i}-\Delta E/2}^{E_{i}+\Delta E/2}
 {\rm d} E'
 \int 
 {\rm d} E_{\nu}
 \frac{{\rm d} \Phi (E_{\nu})}{{\rm d} E_{\nu}}
 P_{\nu_{\mu} \rightarrow \nu_{\tau}}(E_{\nu})
 \sigma_{\rm CC}(E_{\nu})
 R(E_{\nu},E') \epsilon_{\text{eff}},
\end{align}
where ${\rm d} \Phi / {\rm d} E_{\nu}$ is the beam flux,
$\sigma_{\rm CC}$ is the charged current cross section,
$\epsilon_{\text{eff}}$ is the detection efficiency, and
$R$ is the energy smearing function which is assumed to be the
Gaussian distribution,
\begin{align}
R(E_{\nu}, E')
=
 \frac{1}{\sigma(E_{\nu}) \sqrt{2\pi}}
 {\rm e}^{ - \frac{(E_{\nu} - E')^{2} } { 2 \sigma^{2} (E_{\nu}) } },
\end{align}
with $\sigma \equiv 0.15 E_{\nu}$.
$E_{\nu}$ is the neutrino beam energy and $E'$ 
is the reconstructed energy.
The errors for the event normalization $\sigma_{\text{norm}}$
and so-called tilt-error $\sigma_{\text{cal}}$ are
given in the following subsections.
We consider three experimental setups.  

\subsection{NuFACT beam + OPERA-like detector with $L=130$ km}

We consider an OPERA-like detector at a distance of $L=130$ km from 
a Neutrino Factory beam,
which was examined in Ref.~\cite{yasuda}. 
The detector mass is assumed to be 5.0 kton
The matter profile is assumed to be constant with 
the density 2.7 g/cm$^{3}$
although the matter effect itself is not significant in this setup. 

For the signal detection efficiency, the errors, and the backgrounds,
we follow the {\sf glb}-file {\sf OPERA.glb}.
Since this {\sf glb}-file is designed for
the CNGS beam source, the numbers should be modified
for the neutrino factory beam source.
Here, we use the numbers shown in Tab.~\ref{Tab:NuFactOPERA130km}.

\begin{table}[htb]
\begin{tabular}{|c|c|c|c|}
\hline\hline
\multicolumn{2}{|l|}{$\nu_{\mu} \rightarrow \nu_{\tau}$ Appearance}
&$\sigma_{\text{norm}}$
&$\sigma_{\text{cal}}$ \\
\hline
Signal & 0.1056 $\otimes (\nu_{\mu} \rightarrow \nu_{\tau})_{\rm CC}$
 & 0.05 & $10^{-4}$ \\
Background & $3.414 \times 10^{-5} \otimes (\nu_{\mu} \rightarrow
 \nu_{x})_{\rm NC}$ \qquad
 $3.414 \times 10^{-5} \otimes (\bar{\nu}_{e} \rightarrow
 \bar{\nu}_{x})_{\rm NC}$
 & 0.05 & $10^{-4}$ \\
\hline \hline
\end{tabular}
\caption{Rules for the experimental setup NuFACT+OPERA-like detector.}
\label{Tab:NuFactOPERA130km}
\end{table}

\subsection{NuFACT beam + LAr near detector}

In this set up we consider an 
0.1 kt liquid Argon detector at 2 km far away from the
beam source, which has been discussed in Ref.~\cite{ic}.
Here we follow the {\sf glb}-file, {\sf ICARUS.glb} but modify the 
background estimation.
\begin{table}[htb]
\begin{tabular}{|c|c|c|c|}
\hline\hline
\multicolumn{2}{|l|}{$\nu_{\mu} \rightarrow \nu_{\tau}$ Appearance}
&$\sigma_{\text{norm}}$
&$\sigma_{\text{cal}}$ \\
\hline
Signal & 0.0758 $\otimes (\nu_{\mu} \rightarrow \nu_{\tau})_{\rm CC}$
 & 0.05 & $10^{-4}$ \\
Background & $8.502 \times 10^{-5} \otimes (\nu_{\mu} \rightarrow
 \nu_{x})_{\rm NC}$ \qquad
 $8.502 \times 10^{-5} \otimes (\bar{\nu}_{e} \rightarrow
\bar{\nu}_{x})_{\rm NC}$
& 0.05 & $10^{-4}$ \\
\hline \hline
\end{tabular}
\caption{Rules for the experimental setup NuFACT+LAr detector.}
\end{table}

\subsection{NuFACT beam + large LAr far detector}

In order to solve the $\phi$-$(\pi-\phi)$ degeneracy
and the $(\epsilon_{\mu} \epsilon_{\tau})$-$\phi$ {\it quasi-}degeneracy,
it is effective to observe
the matter effect coming from the non-unitarity effect.
To get the matter effect, we need a long baseline.
Here, we set $L=3,000$ km and adopt 3.3 g/cm${}^{3}$ as the matter density.
However,
in such a long baseline setup, we need a huge detector
to collect enough event rates.
We assume 100 kton LAr detector whose rules
are taken from {\sf ICARUS.glb}, which is modified 
as the same manner as the LAr near detector setup.


\end{document}